\documentclass[epj]{svjour}
\usepackage[separate-uncertainty,bracket-numbers = false,
  table-align-uncertainty = true, table-align-exponent=false, 
exponent-product=\cdot,parse-numbers = false]{siunitx}
\usepackage{hyperref}
\usepackage{textcomp}
\usepackage{gensymb}
\usepackage{multirow} 
\usepackage{amsmath}
\catcode`\@=11 
\def\parenbar{\mathpalette\p@renb@r}
\def\p@renb@r#1#2{\vbox{%
		\ifx#1\scriptscriptstyle \dimen@.7em\dimen@ii.2em\else
		\ifx#1\scriptstyle \dimen@.8em\dimen@ii.25em\else
		\dimen@1em\dimen@ii.4em\fi\fi \offinterlineskip
		\ialign{\hfill##\hfill\cr
			\vbox{\hrule width\dimen@ii}\cr
			\noalign{\vskip-.3ex}%
			\hbox to\dimen@{$\mathchar300\hfil\mathchar301$}\cr
			\noalign{\vskip-.3ex}%
			$#1#2$\cr}}}
\catcode`\@=12 
\def\nuan{\ensuremath{\parenbar{\nu}\kern-0.4ex}}

\def\Pipm{\ensuremath{\pi^{\pm}}}
\def\PiP{\ensuremath{\pi^{+}}}
\def\PiM{\ensuremath{\pi^{-}}}

\def\Mu{\ensuremath{\mu^{\pm}}}

\def\MuP{\ensuremath{\mu^{+}}}
\def\Kpm{\ensuremath{K^{\pm}}}

\def\NuANuMu{\ensuremath{\nuan_\mu}}
\def\NuANuE{\ensuremath{\nuan_e}}

\def\NuMu{\ensuremath{\nu_\mu}}

\def\ANuMu{\ensuremath{\bar\nu_\mu}}

\def\PiMuNu{\ensuremath{\Pipm \to \Mu \NuANuMu}}
\def\KMuNu{\ensuremath{\Kpm \to \Mu \NuANuMu}}

\def\nue{\ensuremath{\nu_{e}\ }}
\def\nubare{\ensuremath{\overline{\nu}_{e}}\ }

\def\numu{\ensuremath{\nu_{\mu}\ }}
\def\nubarmu{\ensuremath{\overline{\nu}_{\mu}}}

\newcommand{\deltacp}{\ensuremath{\delta_{CP} \,}}

\newcommand{\fome}{\ensuremath{E^{\rm{vac.}}_{1}}}
\newcommand{\fomem}{\ensuremath{E^{\rm{mat.}}_{1}}}

\newcommand{\EPO}{5.0} 
\newcommand{\ECP}{3.3} 
\newcommand{\ECT}{2.4} 
\newcommand{\ECC}{2.5} 
\newcommand{\EEO}{2.9} 
\newcommand{\EFN}{6.0} 

\usepackage{xcolor,soul}
\usepackage{array}
\usepackage{geometry}
\usepackage{caption}
\usepackage{float}
\usepackage{subcaption}
\usepackage{graphicx}
\usepackage{mathtools}
\usepackage{cuted}

\begin{document}
\title{NuTag: proof-of-concept study for a long-baseline neutrino beam}

\author{A. Baratto-Roldán\inst{1}\thanks{\emph{Corresponding author: A. Baratto-Roldán (CERN). \\ E-mail:anna.baratto.roldan@cern.ch}} \and M. Perrin-Terrin\inst{2} \and E.G. Parozzi\inst{1} \and M.A. Jebramcik\inst{1}\and N. Charitonidis\inst{1}
}                     
%
%
\institute{CERN, BE Department, Esplanade des Particules 1, Meyrin, 1211 Geneva 23, Switzerland \and Aix Marseille Univ, CNRS/IN2P3, CPPM, Marseille, France }


%
\date{Received: date / Revised version: date}
%

\abstract{
The study of neutrino oscillation at accelerators is limited by systematic uncertainties, in particular on the neutrino flux, cross-section, and energy estimates. These systematic uncertainties could be eliminated by a novel experimental technique: neutrino tagging. This technique relies on a new type of neutrino beamline and its associated instrumentation which would enable the kinematical reconstruction of the neutrinos produced in \PiMuNu\ and \KMuNu\ decays.
This article presents a proof-of-concept study for such a tagged beamline, aiming to serve a long baseline neutrino experiment exploiting a megaton scale natural water Cherenkov detector. After optimizing the target and the beamline optics to first order, a complete Monte Carlo simulation of the beamline has been performed. The results show that the beamline provides a meson beam compatible with the operation of the spectrometer, and delivers a neutrino flux sufficient to collect neutrino samples with a size comparable with similar experiments and with other un-tagged long-baseline neutrino experimental proposals. 
\PACS{
      {PACS-key}{discribing text of that key}   \and
      {PACS-key}{discribing text of that key}
     } 
} 

\maketitle
\section{Introduction}
\label{sec:intro}

Neutrino tagging is a novel technique being developed for accelerator-based neutrino experiments~\cite{hep-ph_Perrin-Terrin_2022}. Usually, the characteristics of the neutrinos (energy, flavour, direction, and chirality) are estimated based on the signals they induce when interacting in a neutrino detector. The properties of these interacting neutrinos are then used to infer the overall neutrino flux, which is essential to study neutrino oscillations. However, as only a tiny fraction of neutrinos interact with the detector, the flux calculation is limited by large uncertainties. In addition, the precision on the expected neutrino characteristic is strongly limited by the uncertainties in the models describing the neutrino interactions. 
While the neutrino oscillations studies could so far accommodate these uncertainties, they are becoming the limiting factor for the physics program of the next generations of experiments~\cite{hep-ph_BrancaEtAl_2021,hep-ph_ChatterjeeEtAl_2021,hep-ph_DeRomeriEtAl_2016}.

The neutrino tagging (NuTag) technique~\cite{hep-ph_Perrin-Terrin_2022} proposes to exploit the neutrino production mechanisms to complement the information obtained using the interactions in neutrino detectors. While neutrinos can interact with matter through multiple interaction channels, all exhibiting large experimental signature variability~\cite{hep-ph_SajjadAtharEtAl_2023}, their production mechanisms are dominated by two experimentally well controlled processes: the \PiMuNu\ and the \KMuNu\ decays. Hence, the neutrinos can be kinematically reconstructed from the \Pipm\ and \Mu\ or \Kpm\ and \Mu\ characteristics. These particles being electrically charged, they can be detected with high efficiency and precision~\cite{Knoll}. The neutrinos reconstructed as such are called \textit{tagged neutrinos}.
This reconstruction method offers unmatched precision on the neutrino characteristics. For example, neutrino energy resolutions better than 1\%~\cite{hep-ph_Perrin-Terrin_2022} can be obtained with the tagging, while the typical energy resolutions of neutrino detectors are of the order of 10-30\%~\cite{hep-ph_FriedlandEtAl_2019,hep-ph_HyperKamiokande_2018}. Moreover, as elaborated in \cite{hep-ph_Perrin-Terrin_2022},  the resolutions on the direction and time of arrival of the tagged neutrinos allow associating each interacting neutrino to its corresponding tagged neutrino. As a result, a tagged neutrino experiment can exploit the precise information provided by the tagging to study a wide range of topics both at short and long baseline experiments, such as neutrino interaction models and cross-sections or, neutrino oscillations. The former two are crucial inputs for the upcoming long baseline experiments~\cite{hep-ph_ESPP_2020}, DUNE~\cite{hep-ph_DUNE_2020} and T2HK~\cite{hep-ph_HyperKamiokande_2018}. For the latter, preliminary studies~\cite{hep-ph_Perrin-Terrin_2022} indicate that a tagged neutrino beam with a $\mathcal{O}(\mathrm{Mton})$ scale natural-water Cherenkov detector would reach unmatched precision on the phase controlling the charge-parity violation in the neutrino sector $\delta_\textrm{CP}$.

The concept of neutrino tagging was originally introduced in the 1970s with various experimental proposals~\cite{hep-ph_Pontecorvo_1979,hep-ph_Nedyalkov_1984,hep-ph_Kaftanov_1979,hep-ph_Hand_1969,hep-ph_BoikovEtAl_1980,hep-ph_Bohm_1987,hep-ph_BernsteinEtAl_1990,hep-ph_Bernstein_1989}. Despite their differences, all these designs relied on instrumenting the beamline to detect the meson decay products and, ideally, the mesons themselves. Among these designs, only one has been implemented: the Tagged Neutrino Facility (TNF)~\cite{hep-ph_Bohm_1987} in Serpukhov and it only featured instruments to characterise the mesons decay products. To the authors' best knowledge, the TNF collaboration has not published any results, except preliminary ones showing one neutrino in coincidence with a \MuP~\cite{hep-ph_AnikeevEtAl_1998}.

In a tagged neutrino experiment, the neutrino rate is directly limited by the beam instrumentation performance (e.g. readout capability, time resolution)~\cite{hep-ph_Nedyalkov_1984}. Such limitations probably hindered the development of the tagging technique and led the community to adopt experimental designs with maximum beam intensities towards the maximum number of neutrinos to be detected in the neutrino detectors. 
However, with the  recent progress of silicon pixel technologies, the detection rate capability has strongly increased. The ongoing developments in the silicon pixel detector technology field~\cite{hep-ph_AglieriRinellaEtAl_2019,hep-ph_Lai_2018,hep-ph_ECFA_2020} will allow to reach readout fluxes as high as \SI{100}{MHz/mm^2}, integrated fluences of \SI{10^{16-17}}{n_{eq}/cm^2} and time resolutions as good as \SI{10}{\pico\second}~\cite{hep-ph_Lampis_2022}. In this context, our present work targets the design of a beamline specifically tailored to exploit the advantages of the NuTag technique.

\begin{figure}
\centering
    \begin{subfigure}{0.49\textwidth}
    \includegraphics[width=\textwidth]{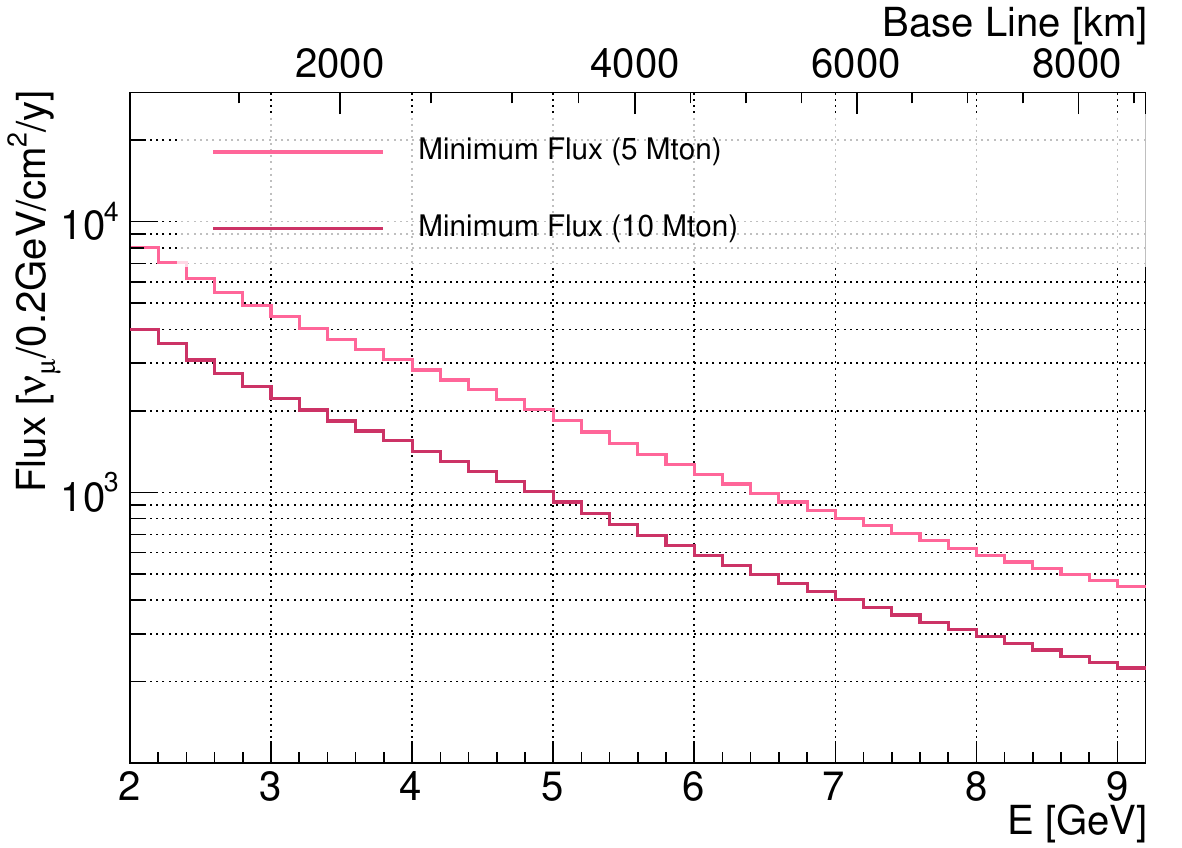}
    \caption{}
    \label{fig-MinFluxNu}
    \end{subfigure} \\
    \hfill
    \begin{subfigure}{0.49\textwidth}
    \includegraphics[width=\textwidth]{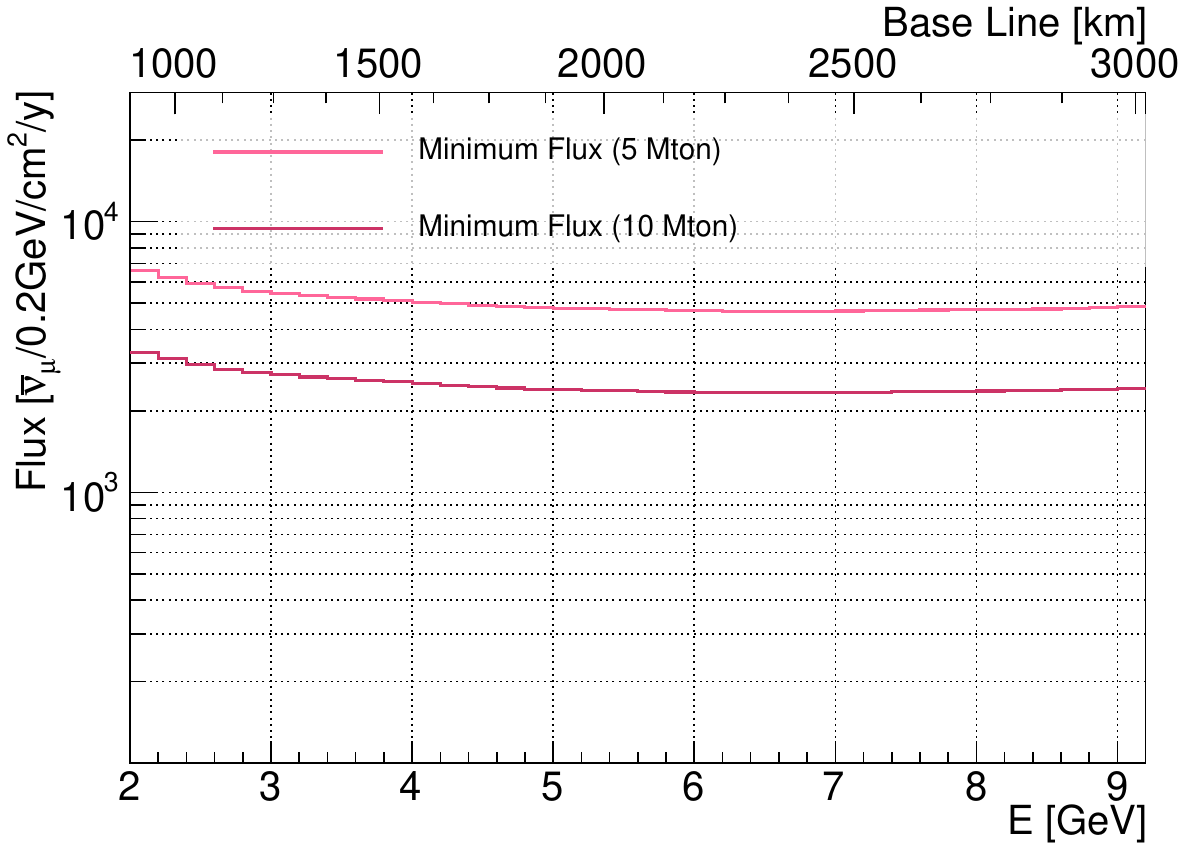}
    \caption{}
    \label{fig-MinFluxANu}
    \end{subfigure}      
\caption{Minimum neutrino (a) and anti-neutrino (b) fluxes required to collect neutrino samples as large as DUNE~\cite{hep-ph_DUNE_2020} assuming a \SI{5}{Mton} and a \SI{10}{Mton} far detectors. The fluxes are shown as a function of the energy of the first oscillation maximum, calculated accounting for matter effects~\cite{hep-ph_MikheyevEtAl_1985,hep-ph_Wolfenstein_1978} with the OscProb software~\cite{OscProb}.}
\label{fig:FluxBound}
\end{figure}

In this article, we present a conceptual design for a tagged long-baseline neutrino beam for the study of neutrino oscillations. In such tagged setups, the neutrino initial flavour, chirality, energy, and direction are determined by the tagger with high precision. Thus, the purpose of the far detector is limited to the identification of the neutrinos flavour. As a result, the granularity of the far detectors is not as crucial as for non-tagged experiments. Hence, an interesting option is to employ $\mathcal{O}(\mathrm{Mton})$ scale natural-water Cherenkov detectors which can attain sizes two orders of magnitude larger than the detectors of the upcoming experiments (DUNE and T2HK). For example, KM3NeT-ORCA~\cite{hep-ph_KM3Net_2016}, under construction off-shore France, will ultimately instrument \SI{6.8}{Mton} of sea water~\cite{hep-ph_KM3NeT_2022}. The energy threshold of this detector is \SI{4}{GeV}~\cite{hep-ph_KM3NeT_2022}. In this paper, we have mainly considered the case study of an experiment with a first oscillation maximum energy (\fome)\footnote{
The \fome is computed for neutrinos in vacuum and using the oscillation parameters values from~\cite{hep-ph_ParticleDataGroup_2022}
} at \SI{\EPO}{GeV}, which corresponds to a baseline length (BL) of \SI{2600}{km} similar to P2O~\cite{hep-ph_AkindinovEtAl_2019}. As the natural water Cherenkov detection technology developed for KM3NeT/ORCA~\cite{hep-ph_KM3NeT_2018} would allow to access both high and low energy ranges, with energies as low as \SI{1}{GeV}~\cite{hep-ph_HofestaedtEtAl_2020}, several other configurations could be envisaged such as:
\begin{itemize}
    \item CERN to Pylos, Greece (BL of \SI{1700}{km} - \fome\ of \SI{\ECP}{GeV})~\cite{hep-ph_BallEtAl_1995},
    \item CERN to Gulf of Taranto, Italy (BL of \SI{1250}{km} - \fome\ of \SI{\ECT}{GeV})~\cite{hep-ph_BallEtAl_2007},
    \item CERN to Capo-Pasero (KM3NeT-ARCA site~\cite{hep-ph_KM3Net_2016}), Italy (BL of \SI{1300}{km} - \fome\ of \SI{\ECC}{GeV}),
    \item European Spallation Source\footnote{Extra accelerating infrastructure would be needed to bring the protons to an energy sufficient to produce neutrino's at the \fome.} to La Seyne-Sur-Mer (KM3NeT-ORCA site~\cite{hep-ph_KM3Net_2016}), France (BL of \SI{1500}{km} -- \fome\ of \SI{\EEO}{GeV}),
    \item Fermilab to Neptune (BL of \SI{3100}{km} - \fome\ of \SI{\EFN}{GeV})~\cite{hep_ph_Vallee}.
\end{itemize}
For this next generation of tagged neutrino experiments to supersede the upcoming ones~\cite{hep-ph_DUNE_2020,hep-ph_HyperKamiokande_2018}, they have to provide, first, a better control of the systematical uncertainties on the neutrino flux and energy estimate and, second, neutrino samples at least as large. While the tagging is expected to drastically reduce theses systematical uncertainties, it will also limit strongly the neutrino beam rate which, together with the detector mass, determines the neutrino sample size. Hence, we use the number of neutrino interactions, $n_{\NuANuE}$,  as a figure of merit for our beam line study. This number can be expressed as
\begin{strip}
\begin{equation}
n_{\NuANuE}(\fomem) = \mathcal{F}_{\NuANuMu}(\fomem) \cdot  P(\NuANuMu\to\NuANuE)(\fomem) \cdot \sigma^{CC}_{\NuANuE} \cdot \mathcal{N}_{\rm{nucl}},
\label{eq:NbNu}
\end{equation}
\end{strip}
\noindent with \fomem\ the energy of the first oscillation maximum derived with the OscProb software~\cite{OscProb} accounting for matter effects~\cite{hep-ph_MikheyevEtAl_1985,hep-ph_Wolfenstein_1978} and using the same oscillation parameters values as~\cite{hep-ph_DUNE_2020}, $\mathcal{F}_{\NuANuMu}$ the neutrino flux, $P(\NuANuMu\to\NuANuE)$ the oscillation probability, $\sigma^{CC}_{\NuANuE}$ the charged current neutrino cross-section and $\mathcal{N}_{nuc}$ the number of nucleons in the detector. The latter is obtained from the detector effective mass, the molar masses of its constitutive elements and the Avogadro number. Hence, using Equation~\eqref{eq:NbNu}, with the number of oscillated \NuANuE\ expected for DUNE at the first oscillation maximum~\cite{hep-ph_DUNE_2020}, one can derive a minimum flux for a given detector mass and \fomem. The minimum fluxes for a 5 and a \SI{10}{Mton} detectors are shown in \autoref{fig:FluxBound}. For neutrinos, both the oscillation probability and cross-section increase with \fomem. Hence, the minimum neutrino flux steadily decreases with \fomem. By contrast, for anti-neutrinos, while the cross-section increases with \fomem, the oscillation probability decreases with it. As a result, the minimum flux is constant above \SI{3}{GeV}. 

\section{Neutrino beamline design general considerations}
\label{sec:design}

The first systematic studies on accelerator neutrino beams were presented in 1965~\cite{Frazinetti}, where the definition of ``conventional" neutrino beams, being used until today, was coined. In conventional neutrino beams, the neutrinos originate in the decay of \Pipm\ or \Kpm\, typically produced by the interaction of a high-intensity, high-momentum proton beam with a target material. The \PiMuNu\ process is dominating over the \KMuNu\  due to the fewer number of kaons produced, and the electron-neutrinos are always subdominant because they are created either by $K^\pm \rightarrow e^\pm \pi^0 \nue (\nubare)$ or by the decay in flight of $\mu^\pm$, like $\pi^+ \rightarrow \mu^+ \numu \rightarrow ( e^+ \nue \nubarmu) \ \numu$. The readers can find an extensive review of the design on accelerator neutrino beams in~\cite{kopp2006} and a more recent discussion in~\cite{app11041644}. For historical reasons~\cite{Dore:2018ldz}, the majority of the accelerator-driven systems for neutrino production are optimized for the maximum neutrino flux rather than for the precision on the knowledge of the neutrino momentum and rate. These $\lq\lq$wide-band'' beams compensate the small interaction cross-sections of the low-GeV/c scale neutrinos with a large flux. 

The production of a wide (or narrow) band neutrino beam can be quite different depending on the physics scope of the experiment being served by it. However, in general, the following considerations must be taken into account in all cases: 

\begin{itemize}
\item{A high-intensity proton beam impinging on a target material to produce the unstable neutrino parent particles is necessary. The primary beam must be of sufficiently high momentum and power so that a satisfactory number of secondary particles, i.e. pions and kaons, are produced~\cite{hep-ph_ParticleDataGroup_2022}. The proton extraction on the target can be fast (of the order of a few \si{\nano\second}), slow (above \SI{700}{\milli\second}) or fast-slow (of the order of \SI{1}{\milli\second} as done in the now dismantled West Area Neutrino Facility~\cite{Acquistapace:286076} at CERN). This may have implications on the target lifetime and may require explicit cooling or special design considerations. If the beam is fastly extracted, as was proposed in the case of the Neutrino Factory~\cite{IDS-NF:2011swj} for instance, then the target design becomes very challenging, and only novel target concepts like W-powder~\cite{PhysRevAccelBeams.21.033401,CAI20222650} or liquid mercury~\cite{RADAIDEH202241} have been proposed for such cases. Even in the case of slowly extracted beams, the radiation damage after an assumed 10 year operation may be an important concern, especially for graphite-based target materials (see for example~\cite{SIMOS2021100028}). } 

\item{Downstream of the target, a focusing section must be installed to transversely select a satisfactory fraction of the produced secondaries at the target. Typically either quadrupole magnets or magnetic horns are being employed, with their respective advantages or disadvantages (discussed extensively in~\cite{app11041644}) and in conjunction with the selected extraction scheme. On top, most of the horns used or proposed worldwide are optimized for focusing a single charge polarity, due to the toroidal field of the horn~\cite{horn1,horn2,horn3}.}

\item{If a momentum selection is necessary for the parent hadrons, or if a central momentum of the pions and therefore the neutrinos is wanted for the experimental scope, a momentum-selection section of the hadron beamline is needed. Typically, this involves bending magnets and collimating slits.}

\item{The overall background imperatively needs to be minimized, in order to avoid the production of secondaries either after the momentum selection section or in the apertures. Indeed, these secondaries will create pions or muons that will subsequently decay to neutrinos of different momentum, possibly reaching also the far detector and generating events that cannot be reconstructed. The largest challenge in this respect is the primary beam, which needs either to be dumped outside the axis of the neutrino trajectory or very early upstream in the line, ideally followed by bending magnets that can remove the particles with wrong momenta. Another argument for the background minimisation is the slow read-out rates of the neutrino detectors, e.g. the reader may refer to the novel NP-04 neutrino detector for the future DUNE experiment, with a readout rate of the order of a few Hz~\cite{Abi_2020}}.

\end{itemize}

\section{The NuTag beamline concept}
\label{sec:beamlineconcept}

\begin{figure*}[!htp]
    \includegraphics[width=\textwidth]{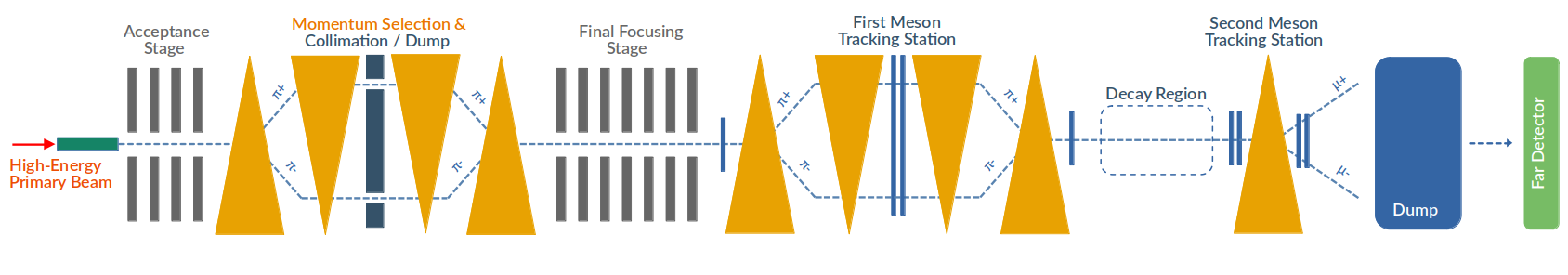}
    \caption{Conceptual scheme of the proposed NuTag beamline. The high energy beam impinging on the target produces secondary particles (only the \PiP\ and \PiM\ beams are shown for simplicity) that are subsequently focused, momentum selected and transported towards the meson tracking station. After their tagging at that point, the secondary beam continues straight towards a decay region where both \PiP\ and \PiM\ will decay to neutrinos. Finally, a second spectrometer measures the momentum, direction and time of arrival of the \Mu\ from the decay, followed by a dump to absorb the remaining surviving beam particles.}
    \label{fig-BeamLineSchematicDiag}
\end{figure*}

A schematic of the proposed beamline concept is shown in \autoref{fig-BeamLineSchematicDiag}. In the proposed beamline, an accelerator delivers a high-energy, high-intensity proton beam to a target. The secondary hadrons emitted are subsequently focused and momentum selected within a very broad momentum range ($\delta{p}/p$ of the order of 25\%), while the primary proton beam is dumped in the thick material between the holes of a collimator that acts as an effective beam dump integrated in this momentum slit. Then, the momentum, direction, and time of arrival of all beam particles are measured by a beam spectrometer made of four dipole magnets, alternated with tracking stations and arranged to form an achromat~\cite{Wiedemann}. These particles then drift in a free space (order of \SI{900}{m}) where the majority will decay. At the end of this drift space, a second spectrometer, made of two pairs of tracking stations installed before and after a large dipole magnet, measures the momentum, direction, and time of arrival of the beam particle decay products, in particular the \Mu's from the \PiMuNu\ and \KMuNu. The information provided by the two spectrometers together allows
to reconstruct the tagged neutrino based on the decays kinematics. The second spectrometer is followed by an absorber that stops the remaining surviving mesons before they decay. The length of the beamline section upstream of the first spectrometer and downstream of the second one should be as short as possible to reduce the number of \Pipm\ and \Kpm\ decaying there, since these cannot be reconstructed\footnote{To estimate the fraction of un-taggable neutrinos, one can conservatively assume that \PiMuNu\ decays occurring upstream of the tagger yield as many neutrinos in the far detector acceptance as the decays occurring downstream of it. Under this assumption, the fraction of un-taggable neutrinos is $( e^{d/\beta\gamma c\tau}-1) / (1-e^{-D/\beta\gamma c\tau}) $ where $d$ and $D$ are the lengths of the beamline sections upstream and downstream of the tagger, $\beta$ and $\gamma$ the reduced velocity and Lorentz boost of the \Pipm, $c$ the speed of light and $\tau$ the \Pipm\ mean lifetime}. Finally, as the neutrino chirality is determined by the charge of the particles in the spectrometer, the hereby proposed NuTag technique does not require removing one of the two beam particle polarities. Hence, the beamline should ideally transport beam particles of both charges. Besides the obvious gain in terms of running time, collecting both neutrino and antineutrino simultaneously is a strong asset to reduce systematical uncertainties for key measurements such as $\delta_\textrm{CP}$.

\subsection{Proposed NuTag beamline layout}
\label{sec:beamlinelayout}

In order to satisfy the physics case discussed in the introduction, the beamline design hereby is optimized for a long-baseline type of facility. The primary beam is considered here to be a slowly extracted, high-momentum proton beam. On one hand, the constraint for the maximum rate that is accepted by the tracking detectors (of the order of \SI{10-100}{MHz/mm^2}) can be only achieved via slow extraction. On the other hand, the high proton momentum a-priori favors the meson production at the momenta of interest ($p<$\SI{20}{\giga\electronvolt}/c), as phenomenologically demonstrated (e.g. the interested reader can observe the production maxima in \cite{hep-ph_ParticleDataGroup_2022} and \cite{Atherton:133786}, where the maximum production for secondaries of momentum $p$ seems to lie within a fraction of 10-20\% of the primary proton momentum). In our case, and using an existing beam example, we chose the \SI{400}{GeV/c} of the CERN Super Proton Synchrotron (SPS). The assumption of the slowly extracted beam also defines essentially the focusing mechanism downstream of the target. In our case, we have based this on large aperture quadrupole magnets. Magnetic horns have not been considered in this study, given the complexity of production and optimisation~\cite{kopp2006}, despite the possible advantages on the acceptance. Furthermore, these devices cannot easily be used in a slow extraction scheme since they don't stand heating beyond a few millisecond. The magnets employed in our design are existing magnets that have been used or are currently in use at CERN experimental areas~\cite{magnetskit}, with known dimensions, properties, and characterized magnetic field gradients. However, for a real implementation, new or different magnets could of course be considered. Concerning the transverse optics design, the first-order matrix approximation is used only as a guide for the beam expected properties. The first-order optics was developed using TRANSPORT~\cite{TRANSPORT} and MADX~\cite{madx}, while the calculation of the hadronic interaction with the target material was performed using the well-known Monte-Carlo software FLUKA~\cite{10.3389/fphy.2021.788253,Boehlen:2109973}. Finally, a custom C++ software was developed to estimate the neutrino flux at the detector based on the meson flux simulated with FLUKA.

\subsubsection{Target and acceptance stage front-end}
\label{sec:target}

The high-intensity, high-momentum beam from an accelerator-driven system, will be slowly extracted on a target to produce the mesons that will then generate the neutrino beam. The dominant production mode being \PiMuNu\, we limited our study to this mode. For this study, a dedicated optimization effort has taken place in order to identify the best target for the described physics case. Various target geometries have been studied; literature research for various other neutrino beams (proposals and existing ones), like CNGS~\cite{Gschwendtner:1053097,Bruno:note138}, NuMi~\cite{jyoti2017study,ADAMSON2016279}, T2K~\cite{ABE2011106} and ENUBET~\cite{Acerbi:2868571}, have demonstrated that graphite is the preferred material, due to its (relatively) low radiological impact and its robustness to high beam powers and temperatures. In our optimization process, we started from lengths and radii values used or proposed in other facilities, combining the two to explore the relative effect of each parameter on the target properties. The different target configurations that demonstrated the best performance in our optimisation study, are summarized in Table~\ref{tab:tgconf}. In our study, all the targets have been simulated in FLUKA as continuous rods of graphite (density \SI{1.7}{g/cm^3}). Considering a right handed Cartesian coordinate system with the beamline aligned along the $z$ axis, the 4-vectors of produced particles in terms of $x$, $x'$, $y$, $y'$, and total momentum $p$ have been scored for all different targets in a perpendicular plane at $z=\SI{0}{\meter}$, the starting point of the beamline, coinciding with the end surface of each target. The primary beam momentum was assumed to be in all cases, \SI{400}{\giga\electronvolt}/c, for reasons discussed in the previous sub-section. 

As discussed in Section~\ref{sec:intro}, different baselines covering different neutrino \fome\ between \SI{\ECT}{\giga\electronvolt} and \SI{\EFN}{\giga\electronvolt} could in principle be envisaged, corresponding roughly to a pion momentum range of $6-\SI{12}{\giga\electronvolt}$/c. The following target studies have been optimized for a reference pion momentum of \SI{12}{\giga\electronvolt}/c (corresponding to a baseline of $\approx\SI{2600}{\kilo\meter}$ with a \fome\ at \SI{\EPO}{\giga\electronvolt}), however, other momenta have been also taken into account, as shown later on.

\begin{table*}
    \centering
    \small 
    \setlength{\tabcolsep}{3pt} 
    \renewcommand{\arraystretch}{1.5} 
    \setlength{\extrarowheight}{2pt} 
    \renewcommand{\arrayrulewidth}{1pt} 
    \begin{tabular}{|c|c|c|>{\centering\arraybackslash}p{7cm}|}
        \hline
        Identifier & Length [cm] & Radius [mm] & Comment \\
        \hline
        A & 70 & 5 & Length of ENUBET proposed target~\cite{Acerbi:2868571} \\
        B & 70 & 13 & Length of ENUBET proposed target~\cite{Acerbi:2868571} with T2K's target radius~\cite{ABE2011106} \\
        C & 126.1 & 2.5 & Length and radius of CNGS target~\cite{Bruno:note138} \\
        D & 126.1 & 13 & Length of CNGS target~\cite{Bruno:note138} with T2K's target radius~\cite{ABE2011106} \\
        E & 94 & 3.7 & Length and radius of NuMi first design target~\cite{ADAMSON2016279} \\
        F & 91.4 & 13 & T2K-like target~\cite{ABE2011106} \\
        \hline
    \end{tabular}
    \caption{Target configurations studied in the framework of the present study.}
    \label{tab:tgconf}
\end{table*}

\begin{figure*}[!htp]
    \includegraphics[width=\textwidth]{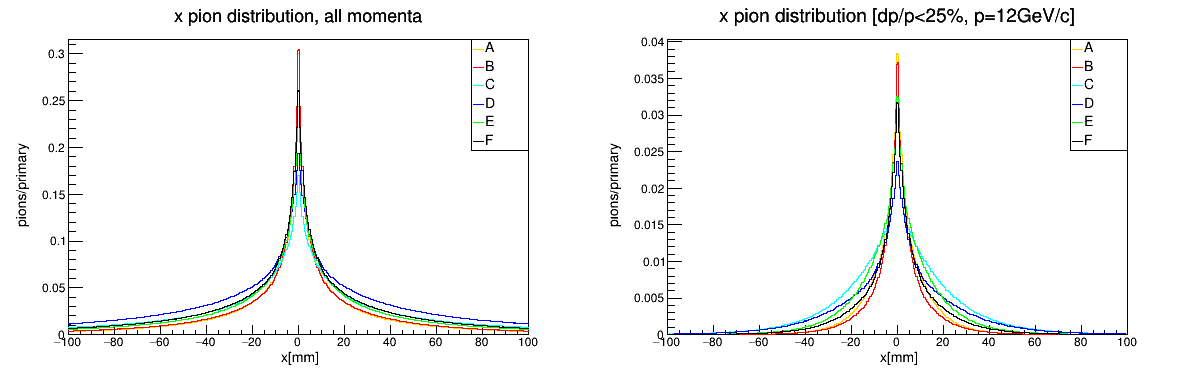}
    \caption{$x$ ($y$) distribution for \Pipm's at $z=0$. On the left, \Pipm's covering the full momentum range; on the right, \Pipm's having a reference central momentum of \SI{12}{\giga\electronvolt}/c assuming a momentum spread of $\pm25\%$}
    \label{fig-targetpositiondist}
\end{figure*}

\begin{figure*}[!htp]
    \includegraphics[width=\textwidth]{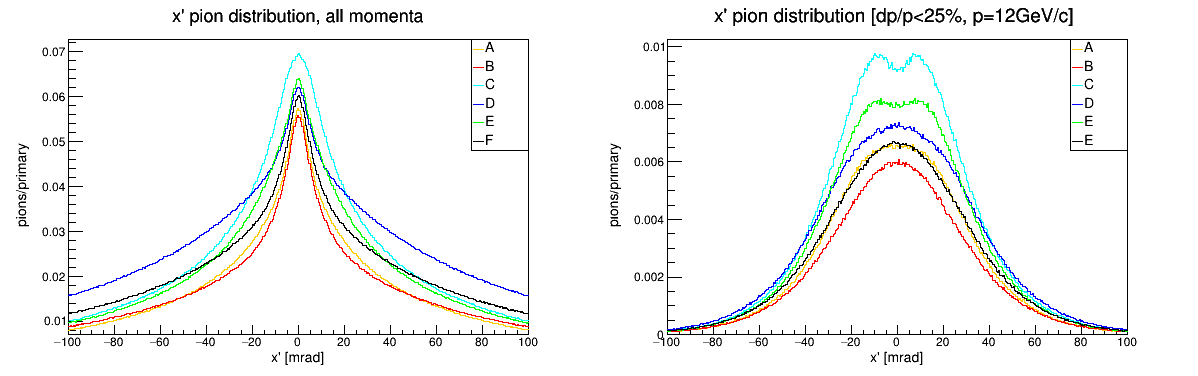}
    \caption{$x'$ ($y'$) distribution  for \Pipm's at $z=0$. On the left, \Pipm's covering the full momentum range; on the right, \Pipm's having a reference central momentum of \SI{12}{\giga\electronvolt}/c assuming a momentum spread of $\pm25\%$}.
    \label{fig-targetangulardist}
\end{figure*}

\begin{figure}
\centering
    \begin{subfigure}{0.49\textwidth}
    \includegraphics[width=\textwidth]{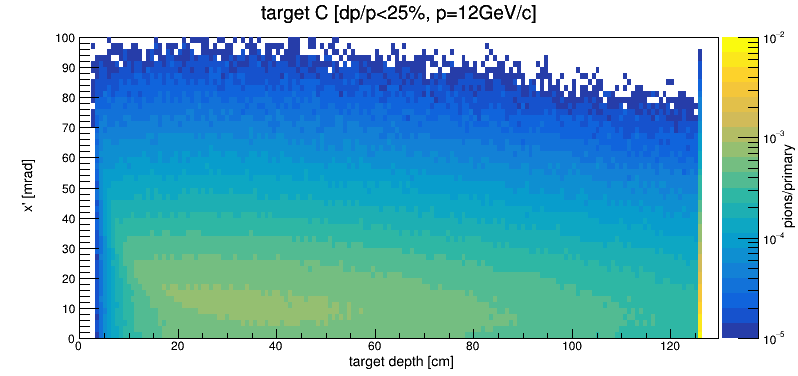}
    \caption{Target C}
    \label{fig-xpvsztargetC}
    \end{subfigure} \\
    \hfill
    \begin{subfigure}{0.49\textwidth}
    \includegraphics[width=\textwidth]{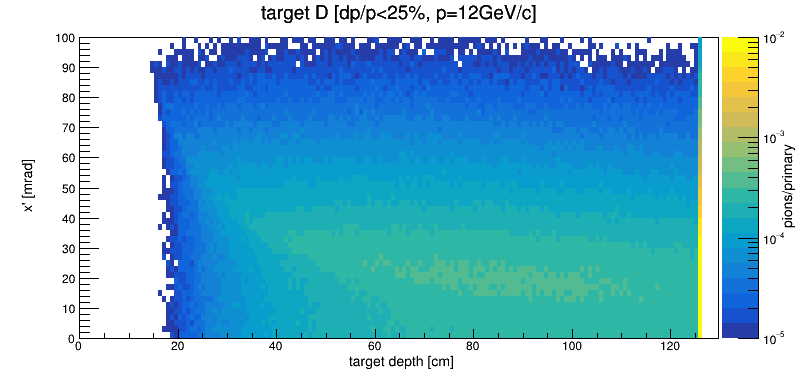}
    \caption{Target D}
    \label{fig-xpvsztargetD}
    \end{subfigure}      
\caption{Angular distribution of \Pipm's in the $\pm25\%$ momentum range around \SI{12}{\giga\electronvolt}/c emerging from the target C (a) and D (b), which share the same length but different radius, as a function of the target depth. The primary proton beam is impinging on the target front face from the left.}
\label{fig-xpvsztarget}
\end{figure}

\autoref{fig-targetpositiondist} and \autoref{fig-targetangulardist} show a comparison, among the various targets, of the $x$ ($y$) and $x'$ ($y'$) distributions for \Pipm's  at $z=0$. In these plots, $x$ ($y$) and $x'$ ($y'$) distributions are shown both for \Pipm's covering the full momentum spectrum and for \Pipm's having a reference central momentum of \SI{12}{\giga\electronvolt}/c, considering a momentum spread of $\pm25\%$. The dip observed in the \Pipm's yield in the bottom right plot of \autoref{fig-targetangulardist}, present in the longer and thinner targets identified as C and E in Table~\ref{tab:tgconf}, is due to re-absorption of the forward emitted \Pipm's ($x'<\arctan(R/L)$ where $R$ is the radius and $L$ the target length) by the target material. This is a combined effect of the target radius and length, as shown in \autoref{fig-xpvsztarget}. This figure represents the \Pipm's angular divergence $x'$ as a function of the target depth for the targets C (\autoref{fig-xpvsztargetC}) and D (\autoref{fig-xpvsztargetD}), which share the same length but different radius. From the colour scale in \autoref{fig-xpvsztargetC} emerges that the \Pipm's in the considered momentum range are largely produced at angles $\arctan(R/L)\approx\SI{2}{\milli\radian}<x'<\SI{20}{\milli\radian}$ at the very beginning of target C, and therefore escape transversely through its sides without being reabsorbed, compared to \Pipm's emitted forward, causing the dip in the distribution on the right of \autoref{fig-targetangulardist}. Being this an effect of the radius to length ratio of the target, the dip in the $x'$ distribution disappears as the target radius gets larger and/or the target length smaller. Indeed, as shown in \autoref{fig-xpvsztargetD}, the larger radius of target D (\SI{13}{mm}) would require much larger emission angles $x'>\arctan(R/L)\approx\SI{10}{\milli\radian}$ for a \Pipm\ to escape transversely through the target sides. Since the probability of emission reduces very quickly for larger angles, the effective length seen by the \Pipm's produced at the beginning of the target is constant and the $x'$ distribution at $z=0$ is (approximately) Gaussian.  

\begin{figure*}[!htp]
    \includegraphics[width=\textwidth]{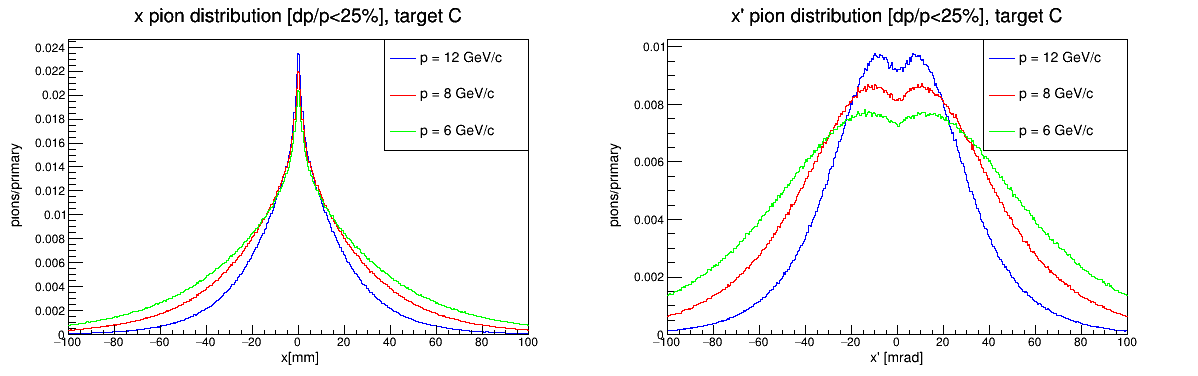}
    \caption{$x$ ($y$) distribution (right) and $x'$ ($y'$) distribution (left) at $z=0$ of \Pipm's in the $\pm25\%$ momentum range around different central momenta. The overall \Pipm\ yield decreases with the central momentum, while a significantly larger amount of \Pipm's is emitted at larger angles.}
    \label{fig-xxpmomentum}
\end{figure*}

Taking into account the results of the aforementioned studies, target C has been chosen as the baseline for the current study. Further target optimization could of course be envisaged, for example by considering spacing the target in different sections with gaps in between to reduce re-absorption, as done for the CNGS~\cite{Bruno:note138} and NuMi~\cite{jyoti2017study} targets. 
Besides our reference $p=\SI{12}{\giga\electronvolt}/c$ \Pipm\ beam, in our study we also considered \Pipm\ beams of \SI{8}{\giga\electronvolt}/c and \SI{6}{\giga\electronvolt}/c. \autoref{fig-xxpmomentum} shows the position and angular distribution of \Pipm's emerging from the target for different central momenta, considering a broad momentum spread of 25\%.

In \autoref{fig-xxpmomentum}, the overall \Pipm\ yield decreases as the central momentum decreases, with a significantly larger amount of \Pipm's emitted at larger angles for the lowest momentum considered. In all cases, the peaks of the distributions are included in the [-20,20]~\si{\milli\radian} angular range. Therefore, a critical aspect of the beamline design is the optimization of the angular acceptance, which needs to be maximized.

\begin{figure*}[!htp]
    \centering
    \includegraphics[width=\textwidth]{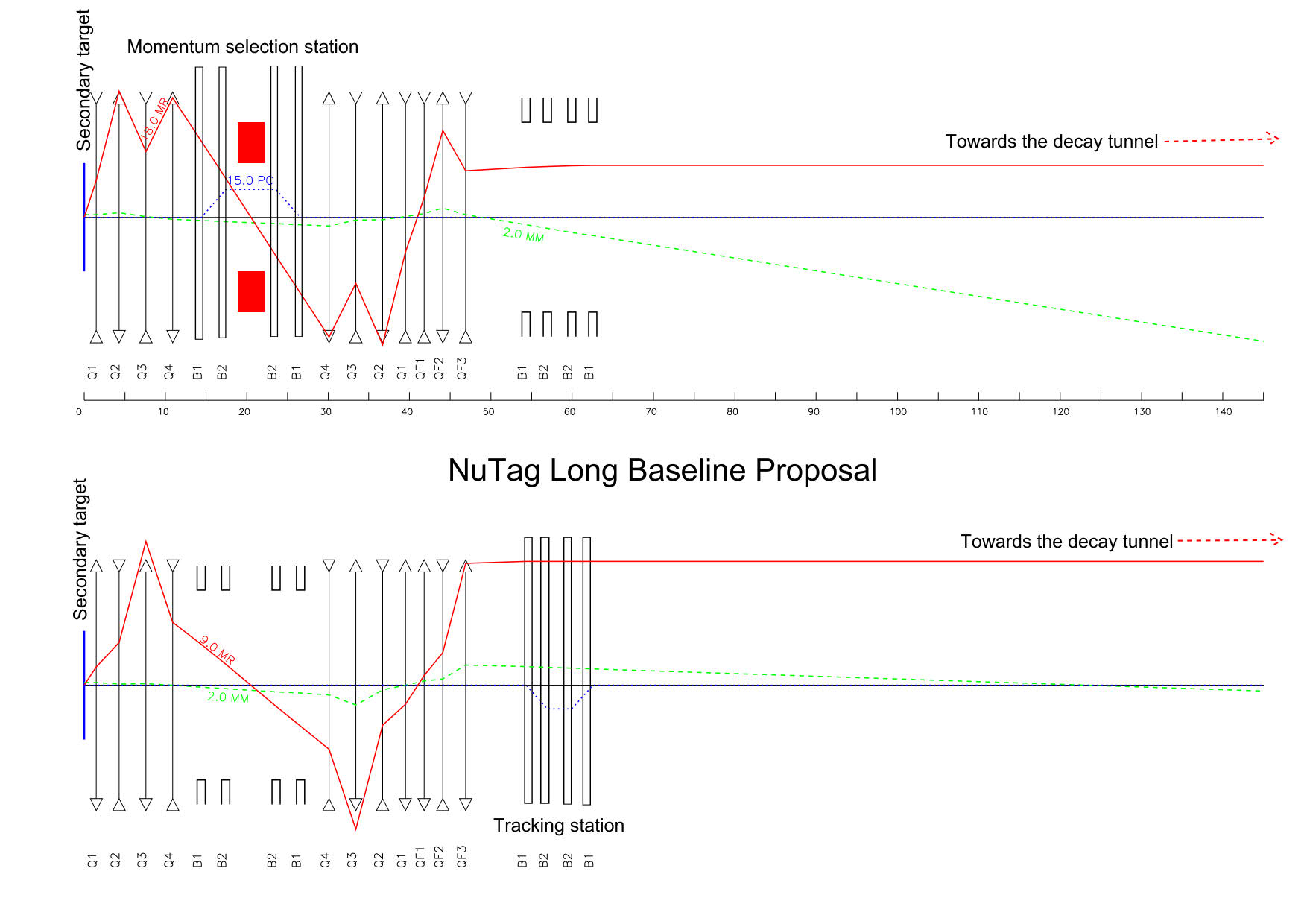}
    \caption{First-order optics of the NuTag proposed beamline. The upper part corresponds to the horizontal plane and the bottom plot to the vertical plane. The red line corresponds to the sine-like ray, the green line to the cosine-like ray, and the blue dotted line to the dispersive ray~\cite{Wiedemann} The tracking detectors are installed in the second achromat, just before the decay region as discussed in the text.}
    \label{fig:opticsv11}
\end{figure*}

\subsubsection{First-order Optics}

\begin{figure}[!htp]
    \centering
    \includegraphics[width=0.49\textwidth]{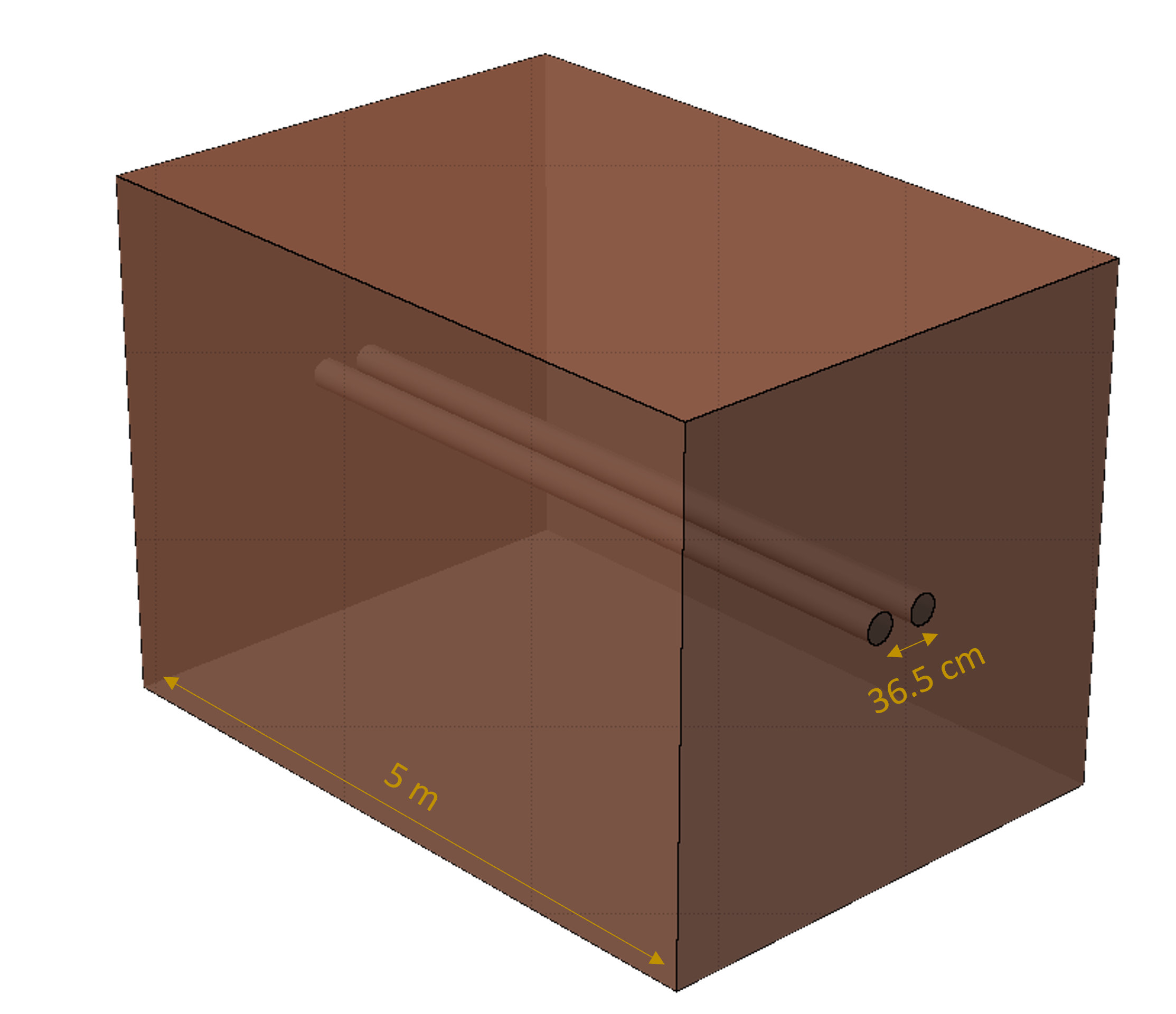}
    \caption{Simplified model (visualized with FLAIR~\cite{FLAIR}) of the custom collimator structure placed in the middle of the first achromat, acting both as a momentum slit for the \PiP\ and \PiM\ beams and as a primary beam dump. It consists of a $\SI{5}{\meter}$ long copper block, with two parallel holes to allow for the passage of the \Pipm\ beams. These holes have a radius of $\SI{10.25}{\centi\meter}$ and a centre to centre distance of $\SI{36.5}{\centi\meter}$.}
    \label{fig-collimator}
\end{figure}

The first-order, transverse optics drawing of the proposed beamline is shown in \autoref{fig:opticsv11}. Downstream of the target, four quadrupoles (notated as \emph{Q1}, \emph{Q2}, \emph{Q3} and \emph{Q4}) ensure the acceptance of a large part of the phase space of particles produced at the target and brings the beam in both the bending and non bending planes, in our case also in both polarities (after an 180$\degree$ phase-advance), at a double focus in the middle of a first 4-bend achromat structure. The quadruplet, apart from allowing a large acceptance in both planes, preserves an absolute symmetry between the R-terms (transport matrix elements, also called  $\mathbf{R}$ matrix) of the horizontal and vertical plane up to the middle of the first achromat. At this exact position, the momentum selection takes place, with the \PiP's and the \PiM's passing through two collimating holes of $\SI{10.25}{\centi\meter}$ radius placed respectively at +$\SI{18.25}{\centi\meter}$ and -$\SI{18.25}{\centi\meter}$ from the centre of a shielding block (see \autoref{fig-collimator}), which also acts as a primary beam dump as mentioned in Section~\ref{sec:beamlineconcept}. 

Downstream of this initial momentum selection, a second quadruplet in series with the first (same quadrupole names correspond to same power converters), brings the momentum-selected beam to 360 degrees phase-advance just before the final focusing stage. At this position, the beam can be cleaned with the use of collimators from background particles that have interacted with the magnet apertures. At this stage, a quadrupole triplet transports (with a 90-degree phase-advance) the double polarity beam towards a second achromat, similar to the first but vertically oriented. The reason for the inverse orientation compared with the upstream achormat was a slight reduction in the acceptance caused by the large spot-size of the beam in the horizontal plane. Also, a vertical orientation favours the placement of shielding in both sides of the tunnel in case this was needed for background reduction. Tracking stations are installed around and inside the second achromat in order to measure the momentum, direction, position and time of all the charged particles, on an event-by-event basis. The large R11 and R33 terms in the horizontal plane along with the \SI{900}{m} length of the decay tunnel, will inevitably result in a transverse beam size of the order of $\mathcal{O}(\si{\meter})$ at the final tracking station, taking also into account the muon decay angles. For this reason, only very large aperture dipoles, such as GOLIATH~\cite{Rosenthal:2310483} or MORPURGO~\cite{MORPURGO1979411}, present today in the North Area at CERN, can serve as spectrometers in the final tracking station, the details of which lie beyond the scope of this paper.  

The described optics predict an angular acceptance of $\mathcal{O}$(20)\si{\milli\radian} in the horizontal and $\mathcal{O}$(9)\si{\milli\radian} in the vertical plane, as will be discussed later in Section~\ref{sec:res}. 

As already mentioned, the part of the primary beam that does not interact with the target ($<<1\%$), is dumped on the large collimator structure (shown in \autoref{fig-collimator}) between the two bending dipoles of the first achromat. Indeed, a sufficient separation between the \PiP\ and \PiM\ beams (roughly $\SI{27}{\centi\meter}$ for the central momentum) ensures that the $\Pipm$ beams pass through the holes and continue downstream, while the protons and all the neutral or off-momentum charged particles are dumped. The collimator/dump structure proposed here is purely conceptual, serving as first assumption, and a proper engineering design would have to follow in future studies.  

\subsection{Meson tracking station}
\label{sec:tracker}

The beam spectrometer has the purpose of measuring the time, position, direction, charge and momentum of all charged beam particles. The detector design is conceptually similar to the NA62 GigaTraKer (GTK)~\cite{hep-ph_AglieriRinellaEtAl_2019,hep-ph_NA62_2017} and its proposed upgrade for the HIKE experiment~\cite{hep-ph_HIKE_2023}. The spectrometer is placed at the second achromat and is composed of a set of tracking stations made from silicon pixel detectors which are located before, after and inside this achromat. The station should be as thin as possible to reduce the multiple coulomb scattering of particles. A material budget of 0.5\% X$_0$ per station was achieved for the NA62 GTK.

The measurements from all the stations allow to reconstruct the trajectory and momentum of each beam charged particle traversing the achromat. Indeed, apart from scattering effects, the trajectories of the particles before and after the achromat are aligned and parallel to the trajectory described by the particle in the innermost drift space of the achromat. The distance between these parallel trajectories is inversely proportional to the particle momentum. As for the NA62 GTK~\cite{hep-ph_AglieriRinellaEtAl_2019}, the material budget of the station is expected to be one of the dominant factors limiting the precision to which the particle's direction and momentum can be determined.

For a configuration with three tracking planes, placed before, after and inside the achromat, as the initial GTK design, the momentum resolution can be expressed as~\cite{hep-ph_NA62_2010}:
\begin{strip}
\begin{equation}
\frac{\delta p}{p} = \frac{1}{\Delta} \cdot \sqrt{
\sigma^2_y +
\left(\frac{\sigma_y\cdot d_{12}}{d_{12}+d_{23}}\right)^2  +
\left(\frac{\sigma_y\cdot d_{23}}{d_{12}+d_{23}}\right)^2  +
\left(\frac{\delta_{\rm{MCS}}\cdot d_{13}\cdot d_{23}}{d_{12}+d_{23}}\right)^2  
} \, ,
\end{equation}
\end{strip}
\noindent where:

\begin{itemize}
    \item $\Delta$ is the distance between the trajectories inside and outside the achromat for the nominal momentum and is determined by the dipole magnet bending angle,
    \item $\delta_{12}$ and $\delta_{23}$ are the distances along the $z$-axis between the first and second and, second and third tracking planes,
    \item $\sigma_y$ is the resolution on the $y$ coordinate of the particle at the tracking planes and is proportional to the pixel size,
    \item $\delta_{\rm{MCS}}$ is the mean scattering angle due to the multiple Coulomb scattering in the tracking plane.
\end{itemize}  

\begin{figure}[!htp]
    \centering
    \includegraphics[width=0.49\textwidth]{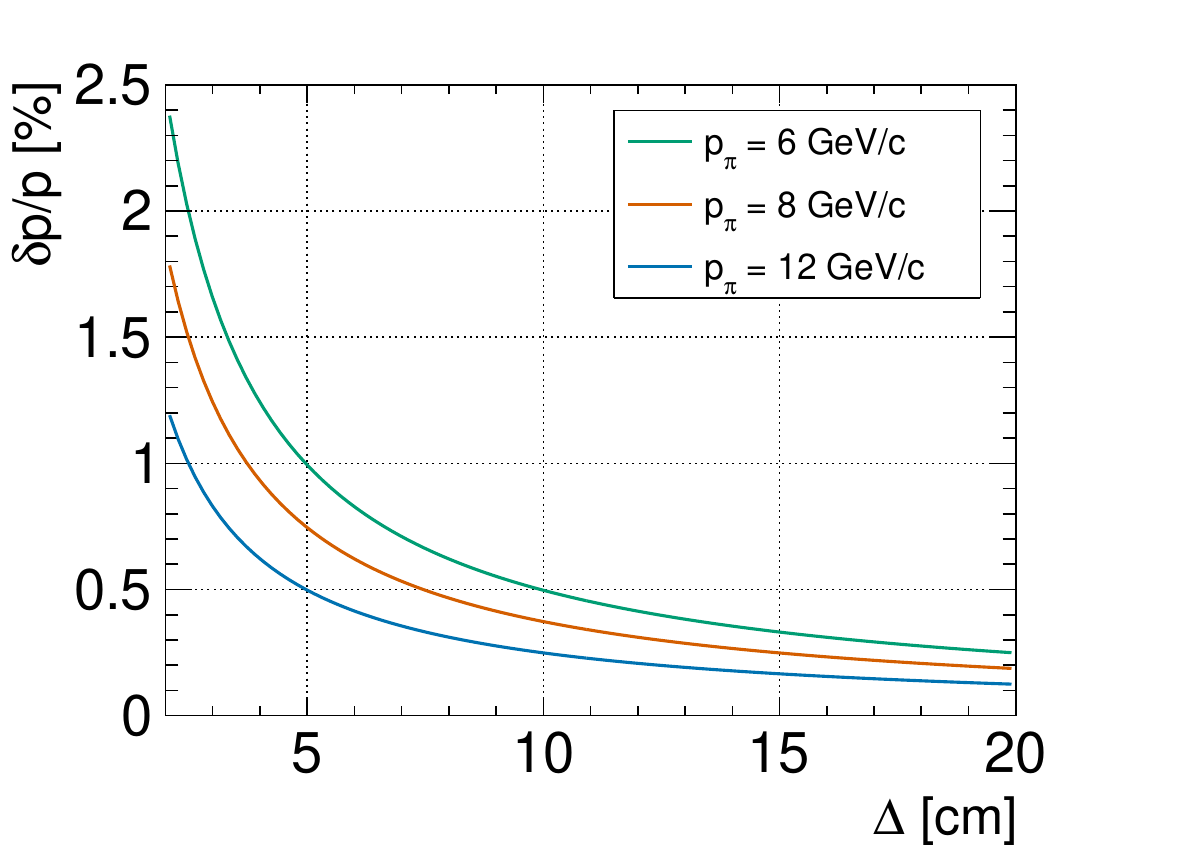}
    \caption{Momentum resolution on the \PiP\ provided by the meson spectrometer, assuming \SI{6}{m} spacing between consecutive tracking stations, a pixel size of \SI{45}{\micro\meter} and a tracking plane thickness equivalent to 0.5\% X$_0$.}
    \label{fig-MomentumReso}
\end{figure}

The momentum resolution is shown as a function of $\Delta$ in \autoref{fig-MomentumReso}, assuming \SI{6}{m} spacing between consecutive tracking stations, a pixel size of \SI{45}{\micro\meter} and a tracking plane thickness equivalent to 0.5\% X$_0$. In the proposed beamline, $\Delta = \SI{11.7}{\centi\meter}$ for a bending angle of \SI{45}{\milli\radian}, which corresponds to a $\delta p/p$ of 0.21\% for a \SI{12}{GeV/c} pion beam. The \Pipm\ momentum resolution provides a good estimate of the neutrino energy resolution for a LBL setup, as the energy of a neutrino emitted collinear to the \Pipm\ is $0.43\cdot E_{\pi}$, where $E_{\pi}$ is the \Pipm\ energy. Hence, as long as $\Delta$ is larger than \SI{5}{cm}, a tagged neutrino experiment is able to provide a neutrino energy resolution better than 1\%.

The pixel technology considered for the station is similar to the one envisaged for the high luminosity phase of the LHC experiments~\cite{LHCbU2_TDR_2022} and is expected to be able to operate at a particle flux as high as \SI{10-100}{MHz/mm^2}~\cite{hep-ph_Lai_2018} and at an integrated fluence of up to \SI{10^{16-17}}{n_{eq}/cm^2}.
The former specification sets a strong constraint on the peak flux of charged beam particles at the second achromat. In addition, the beam transverse size should remain \SI{<0.1}{m^2} as assembling larger pixel modules with a material budget of 0.5\% X$_0$ (i.e. about \SI{500}{\micro\meter} of silicon) would be mechanically challenging. 

\section{Simulated performance}
\label{sec:res}

\begin{figure*}[!htp]
    \centering
    \includegraphics[width=\textwidth]{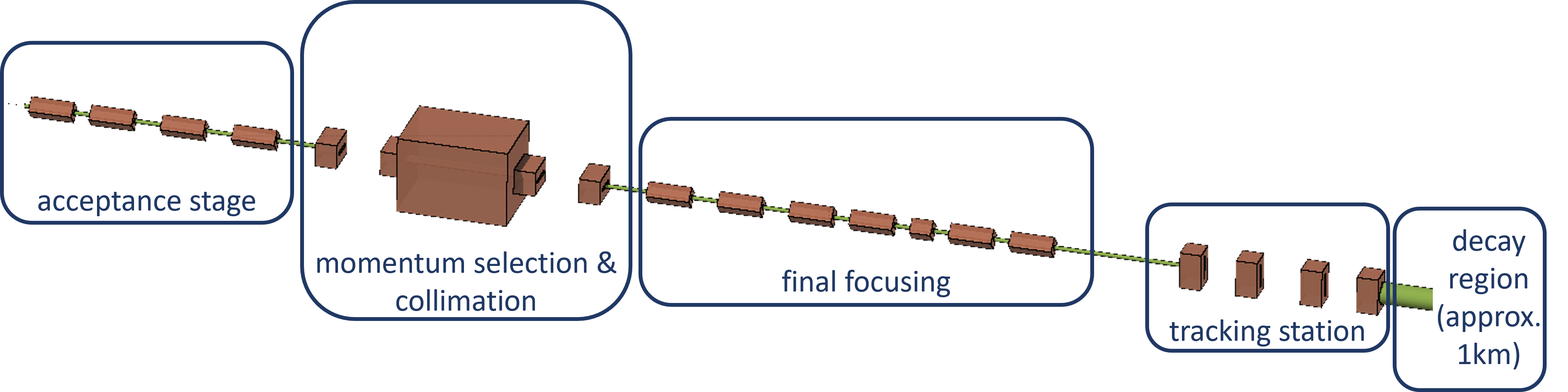}
    \caption{FLUKA model of the proposed beamline, visualized with FLAIR. The acceptance stage is followed by the collimation and momentum selection station, while afterwards the beam is transported to the downstream tracking station. The modelling of the second meson tracking station and dump shown in \autoref{fig-BeamLineSchematicDiag} is beyond the scope of this study.}
    \label{fig-FLUKAmodel}
\end{figure*}

In order to validate the operational principle of the line, a full FLUKA model has been developed up to the second achromat; its geometrical layout is shown in \autoref{fig-FLUKAmodel}. This model includes a complete description of the magnetic two-dimensional field map (both in the gap and in the yoke) for each quadrupole present in the beamline, since the magnets used for this study (QPL and QPS~\cite{magnetskit}) are well characterized and currently in use in many secondary beamlines at CERN. For the bending magnets however, a uniform field throughout the aperture and with no fields in the yoke is assumed. To transport both \PiP\ and \PiM\ beams with large angular acceptance and ensuring a satisfactory momentum selection, bending magnets with larger horizontal apertures than the ones currently present at CERN would be necessary. These magnets, would be identical to the existing MBPS magnets in use today (and described also in \cite{magnetskit}) just with a larger horizontal aperture of 600 mm. The characteristics of the proposed magnets are summarized in Table~\ref{tab:magnets}.

\begin{table*}
    \centering
    \small 
    \setlength{\tabcolsep}{3pt} 
    \renewcommand{\arraystretch}{1.5} 
    \setlength{\extrarowheight}{2pt} 
    \renewcommand{\arrayrulewidth}{1pt} 
\begin{tabular}{|>{\centering\arraybackslash}p{1.5cm}|>{\centering\arraybackslash}p{2.0cm}|>{\centering\arraybackslash}p{1.2cm}|>{\centering\arraybackslash}p{1.7cm}|>{\centering\arraybackslash}p{1.7cm}|>{\centering\arraybackslash}p{1.7cm}|>{\centering\arraybackslash}p{1.7cm}|>{\centering\arraybackslash}p{1.7cm}|}
    \hline
    Name & Type & Length [m] & Aperture type & Aperture Radius [mm] & Aperture Width [mm] & Aperture Height [mm] & Peak Field (Gradient) [T] ([T/m]) \\
    \hline
    QPS & Quadrupole & 1 & Circular & 100 & --- & --- & 10.5 \\
    QPL & Quadrupole & 2 & Circular & 100 & --- & --- & 10.5 \\
    MPBS & Dipole & 1 & Rectan. & --- & 300/600 & 140 & 1.84 \\
    \hline
\end{tabular}

    \caption{Specifications and characteristics of the magnet types used in this study~\cite{magnetskit}. A possible adaptation of the MBPS bending magnet, with a larger aperture width ($\SI{600}{\milli\meter}$ compared to the standard $\SI{300}{\milli\meter}$), has been considered in the FLUKA simulations, to allow for the simultaneous transportation of both pion polarities.}
    \label{tab:magnets}
\end{table*}

In our study, we employed a simple and robust methodology in order to (a) precisely calculate the neutrinos expected in the far detector while (b) avoiding performing computationally expensive Monte-Carlo simulations and/or variance reduction techniques that could create extra uncertainties in the final yield. Specifically, we followed the steps listed below: 

\begin{enumerate}
\item Following the target optimisation described in Section~\ref{sec:target}, we subsequently validated the hadron beam optics and the resulting acceptance based on analytic calculations. At this stage, we made sure that no elements are overly restricting the acceptance and we confirmed the robustness of our computational framework (see Section~\ref{sec:acceptancecalculation}). 
\item Subsequently, we developed a detailed FLUKA model for the beamline and used the produced particle phase space, based on the analysis in Section~\ref{sec:acceptancecalculation} in order to calculate the expected transmission through the beamline (see Section~\ref{sec:flukatransmission}). We validated the overall beamline behaviour by running a full FLUKA simulation of the primary beam, the target and all the magnetic elements down to the beginning of the decay region.
\item Using the results from the step above, we performed analytic kinematic calculations for the expected number and momenta of neutrinos that will reach the far detector (see Section~\ref{sec:piondecaykin}). 
\end{enumerate}

\subsection{Linear acceptance calculation}
\label{sec:acceptancecalculation}

In order to validate the calculated optics, as a first step, we evaluated the phase space of accepted \Pipm's that can be transported from the target to the end of the beamline. This acceptance depends strongly on the aperture limitations of the line. These limitations are defined by the magnets apertures, the collimator holes and the beam pipes. The beam size and angular distribution in each element of the line is defined by the transport matrix, that can be easily retrieved via any optics code or analytical calculations. Therefore, by projecting backwards the individual apertures and mapping the calculation to the beginning of the line, it is possible to identify the smallest limiting aperture and thus the fraction of pions emitted from the target that would eventually be transported to the end of the line. 

\begin{figure*}[!htp]
    \centering
    \begin{subfigure}{0.95\textwidth}
    \centering
    \includegraphics[width=0.95\textwidth]{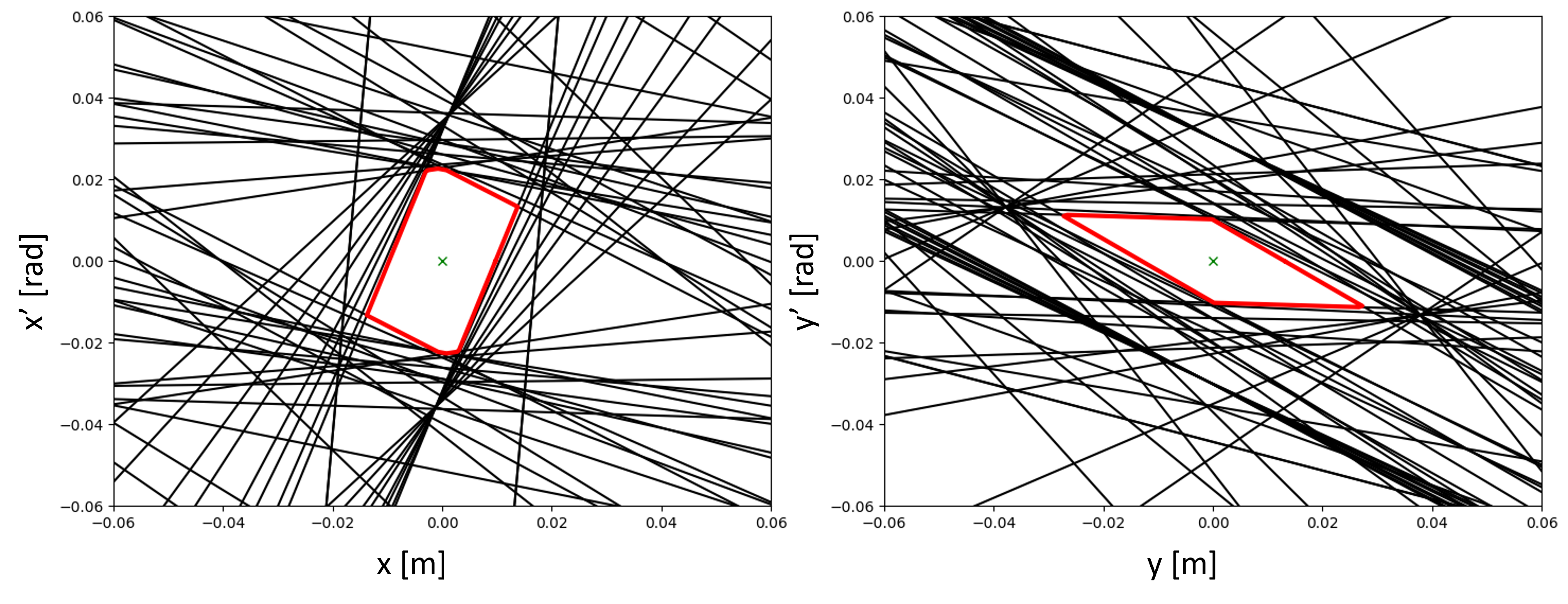}
    \caption{}
    \label{fig-phasespacepionplus}
    \end{subfigure} \\
    \begin{subfigure}{0.95\textwidth}
    \centering
    \includegraphics[width=0.95\textwidth]{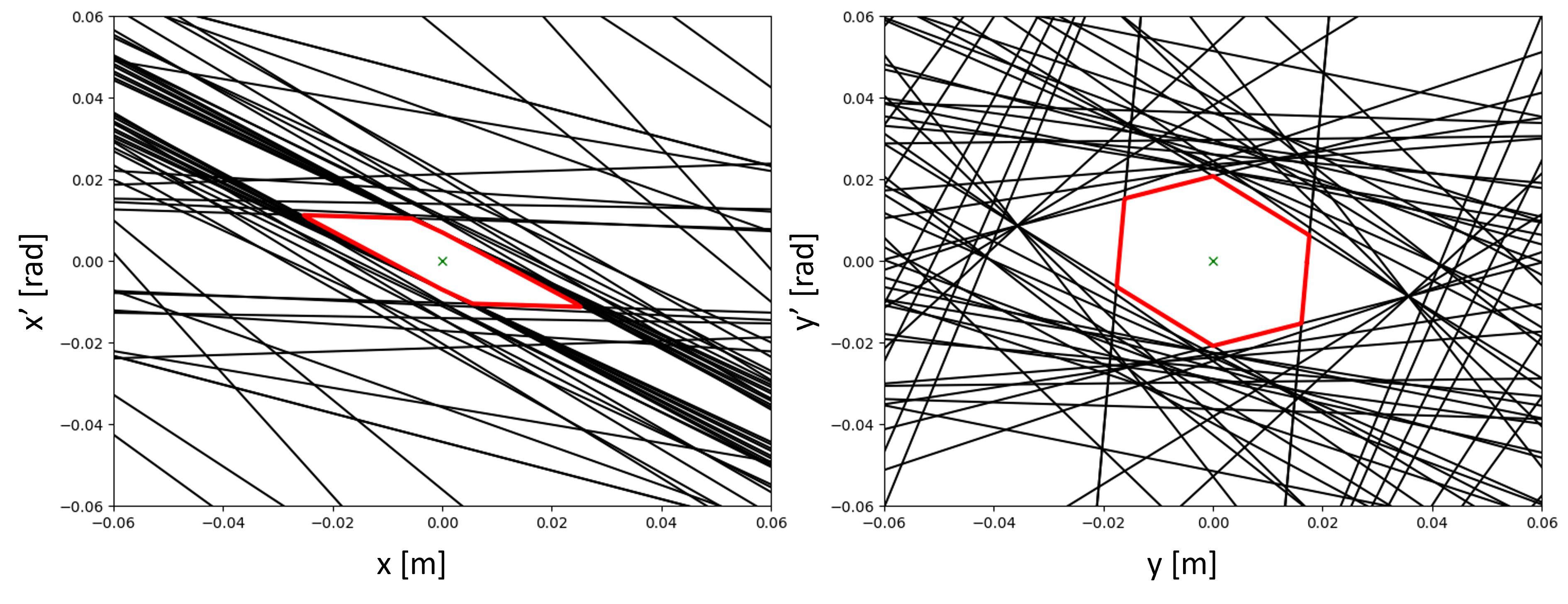}
    \caption{}
    \label{fig-phasespacepionminus}
    \end{subfigure}
    \caption{Accepted phase space of the nominal momentum \PiP\ (a) and \PiM\ (b) beam: (left) horizontal plane and (right) vertical plane. Each of the black lines corresponds to an aperture $\lq\lq$wall" or limitation present along the beamline at some point $z$, projected backwards to the start of the line. The red polygon represents the remaining accepted phase space, obtained by connecting the intersection points of the most inner aperture walls. Compared to \PiP, the \PiM\ $y'$ acceptance is strongly constrained by the vertical gap of the first and last bending magnets in the first achromat.}
    \label{fig-phasespacepions}
\end{figure*}

\begin{figure*}[!t]
    \centering
    \begin{subfigure}{0.95\textwidth}
    \centering
    \includegraphics[width=0.95\textwidth]{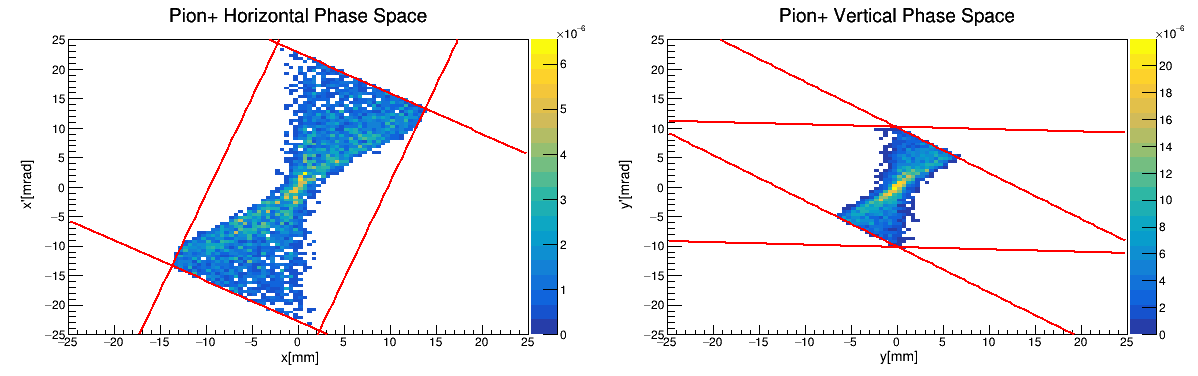}
    \caption{}
    \label{fig-phasespacepionplusfluka}
    \end{subfigure} \\
    \begin{subfigure}{0.95\textwidth}
    \centering
    \includegraphics[width=0.95\textwidth]{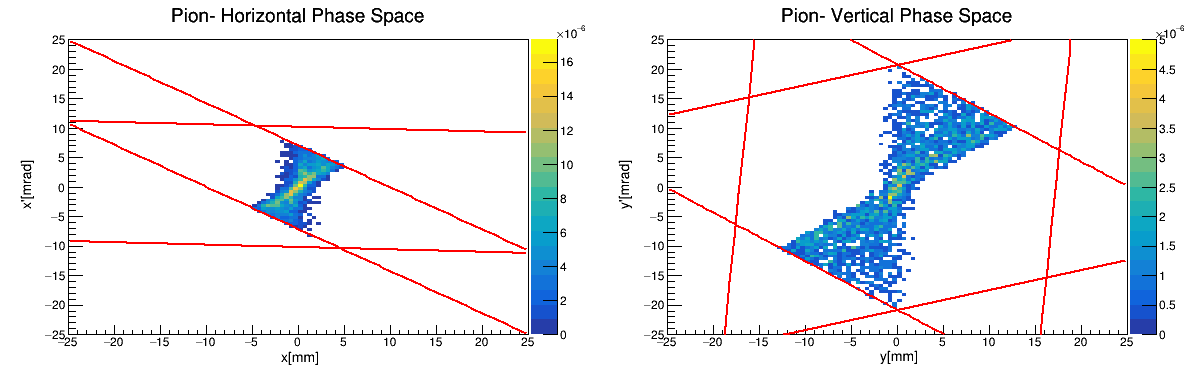}
    \caption{}
    \label{fig-phasespacepionminusfluka}
    \end{subfigure}
    \caption{Target sample \PiP\ (a) and \PiM\ (b) accepted phase space in the $\pm1\%$ momentum range around \SI{12}{\giga\electronvolt}/c. The selection criteria identified in Section~\ref{sec:acceptancecalculation} have been applied to the \Pipm\ yields emitted by the target, obtained from FLUKA simulations, to identify the phase space theoretically transported through the line. Results are normalized per primary proton. As expected by the linear optics, the horizontal plane of the \PiM\ is very similar with the vertical plane of the \PiP\ and vice-versa, as discussed in the text.}
    \label{fig-phasespacepionsfluka}
\end{figure*}

\autoref{fig-phasespacepionplus} and \autoref{fig-phasespacepionminus} show the accepted phase space for the nominal momentum in the horizontal and vertical planes for \PiP\ and \PiM\, respectively. As anticipated from the first-order optics calculation shown in \autoref{fig:opticsv11}, the maximum horizontal and vertical angular acceptances are $\sim$$\SI{20}{\milli\radian}$ and $\sim$$\SI{9}{\milli\radian}$ for \PiP\ and inverted for \PiM. On the other hand, the horizontal and vertical position acceptances are $\sim$$\SI{15}{\mm}$ and $\sim$$\SI{25}{\mm}$ for \PiP\ and the opposite for \PiM. The differences observed in the \Pipm\ accepted phase spaces, already at the first order optics level, are due to the asymmetric aperture of the bending magnets that constitute the two achromats. Indeed, the small vertical gap (140 mm) of the MBPS magnets constrains strongly the $y'$ acceptance for the \PiM\ beam, while the acceptance of the \PiP\ is constrained by the majority of elements present in the line, in both planes. 

Particles with momenta different from the nominal momentum, however, are subject to  chromatic aberrations that become particularly important for large momentum offsets. For momentum offsets beyond $\left| \delta p/p_0 \right| > 10\%$, the limit of the applicability of linear optics is reached, since the first-order approximation (used in the TWISS module of MAD-X used to calculate the optics) is no longer valid~\cite{brown}. For a more accurate evaluation of these particles trajectories, a Monte Carlo simulation was used as will be shown in Section~\ref{sec:flukatransmission}.

\subsection{FLUKA simulations and beamline transmission}
\label{sec:flukatransmission}

The acceptance selection criteria depicted by the red lines in \autoref{fig-phasespacepionplus} and \autoref{fig-phasespacepionminus} have been applied to the \Pipm\ yields produced by the target simulations and scored in FLUKA, as described in Section~\ref{sec:target}. The results of this selection are shown in \autoref{fig-phasespacepionsfluka} for \Pipm\ in a $\pm1\%$ momentum range around \SI{12}{\giga\electronvolt}/c, to stay within the validity limits of the first order optics and avoid second order chromatic effects. 

\begin{figure}[!htp]
    \centering
    \includegraphics[width=0.49\textwidth]{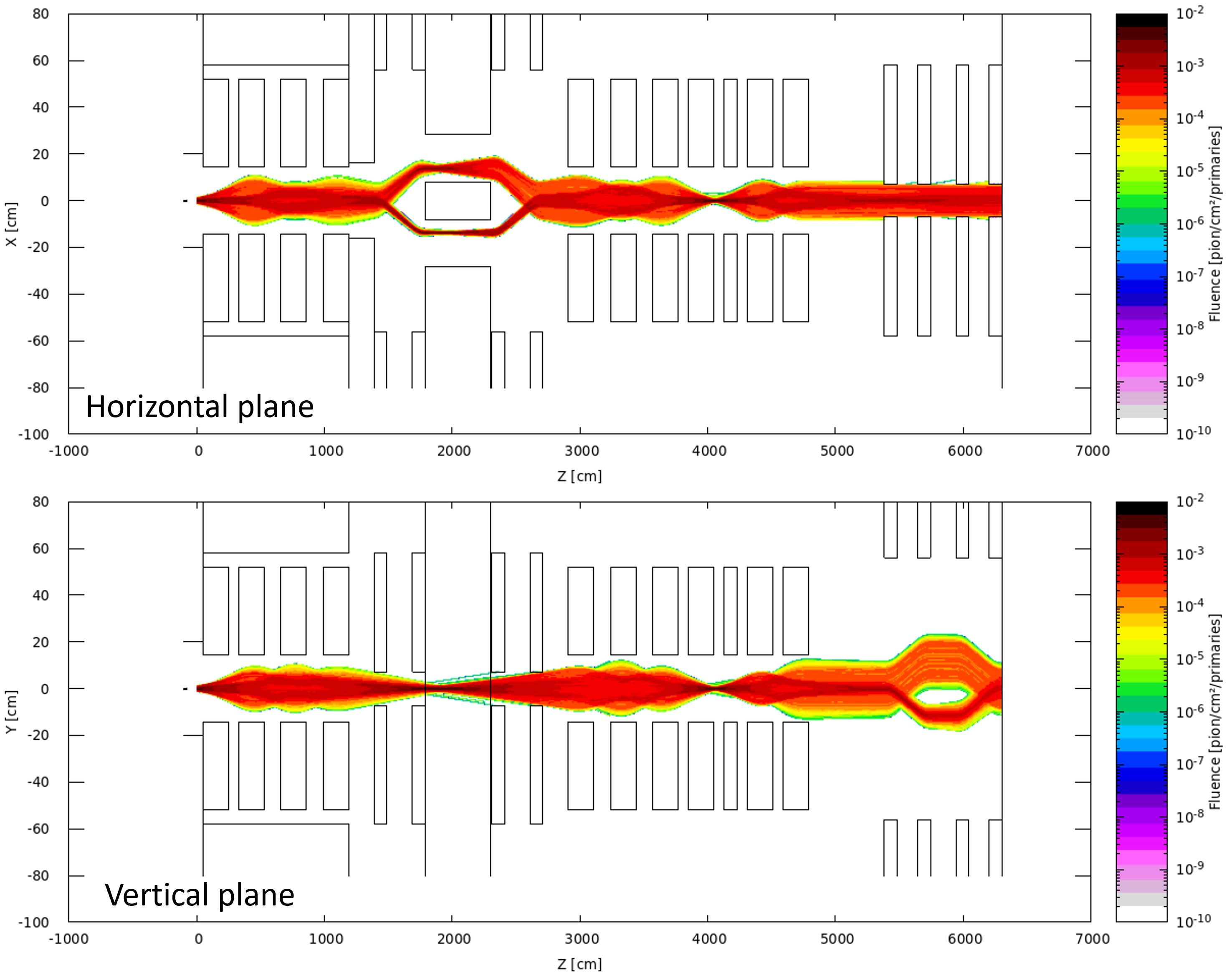}
    \caption{FLUKA simulation of accepted \PiP\ and \PiM\ phase spaces identified in \autoref{fig-phasespacepionsfluka} transported through the beamline. (Top) horizontal and (Bottom) vertical planes. All the results are normalized per primary proton. The trajectories of the particles inside the decay region are not shown.}
    \label{fig-pions1dpp}
\end{figure}

Starting from the identified accepted phase space, and assuming a 1\% momentum acceptance, we performed extensive FLUKA simulations to validate the transmission of the momentum-selected hadron beams through the beamline (i.e., excluding all background), modelling all elements outside the beam-pipe as the FLUKA ``blackhole" i.e, a fully absorbent material that prevents any further transportation of interacting particles. The result of these FLUKA simulations is shown in \autoref{fig-pions1dpp}. It can be seen that for this small momentum spread, the beam trajectory is in very good agreement with the linear optics for both \PiP\ and \PiM. At the momentum-selection position, the slight astigmatism observed between the horizontal and vertical plane, is due to the slightly different magnification terms (green dashed lines in \autoref{fig:opticsv11}) along with the dispersion present in the horizontal plane. The result is that the beam waist is displaced by a few cm upstream. 

\subsection{Full Simulation of the beamline and calculation of momentum acceptance}

\begin{figure*}[!htp]
    \centering
    \includegraphics[width=0.9\textwidth]{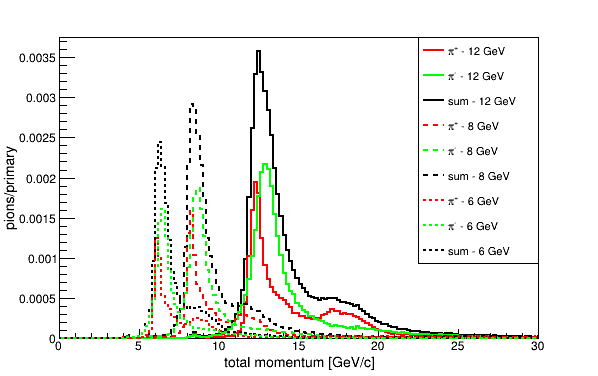}
    \caption{Momentum spectrum of \Pipm's at the start of the decay region, after having been transported through the beamline. The spectrum is shown for three different beamline momenta: $\SI{12}{\giga\electronvolt}$ (solid line), $\SI{8}{\giga\electronvolt}$ (dashed line) and $\SI{6}{\giga\electronvolt}$ (dotted line). The individual contribution to the full spectrum (black) of \PiP (red) and \PiM beams is also shown for each reference momentum.}
    \label{fig-spectrum}
\end{figure*}

To obtain a fully realistic estimation of the number of the \Pipm's transmitted through the line and their momentum distribution (that defines the ``momentum acceptance`` of the beamline), a second simulation was performed using the whole target sample as a source. 
From this simulation, we derived the momentum distribution of the transmitted pion beams that we show in \autoref{fig-spectrum} for three different reference momenta. It can be seen that the centroid of the \PiP\ and \PiM\ distributions do not coincide exactly, being the first peaked around $\SI{12}{\giga\electronvolt}$/c and the second around $\SI{13}{\giga\electronvolt}$/c for a $\SI{12}{\giga\electronvolt}$/c reference momentum. This effect can be explained by second order aberrations and is independent on the reference momentum chosen, since the two beams \PiP\ and the \PiM\ have waists in different positions along z, compared to the focus position, and therefore are (slightly) differently momentum-selected. 

\begin{figure}[!htp]
    \centering
    \includegraphics[width=0.49\textwidth]{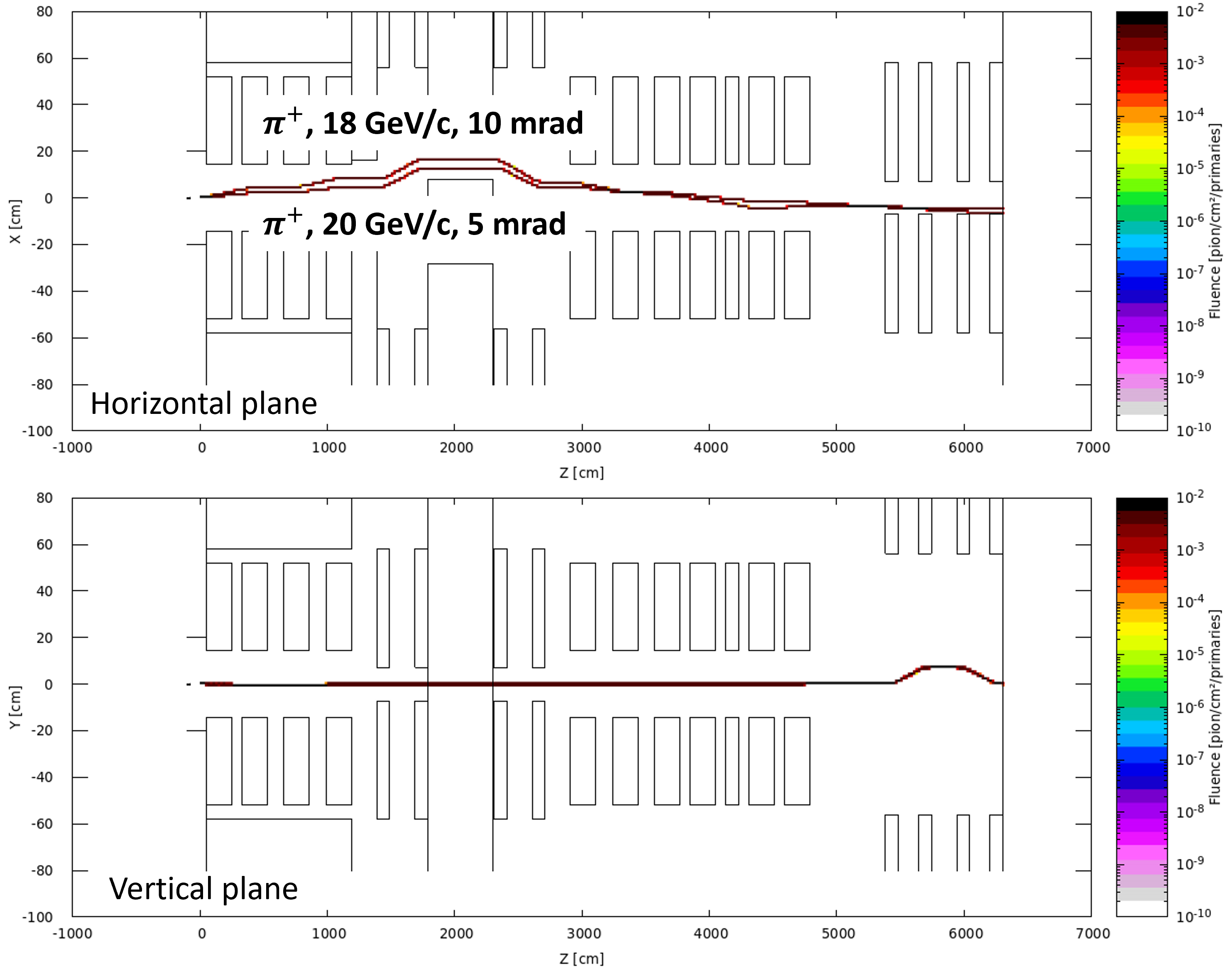}
    \caption{Two example trajectories of higher momentum \PiP's, $\SI{18}{\giga\electronvolt}$/c and $\SI{20}{\giga\electronvolt}$/c, emitted at large angles from the target, $\SI{10}{\milli\radian}$/c and $\SI{5}{\milli\radian}$/c, respectively, and their transmission through the beamline. These larger momentum, but large offset particles constitute the species that populate the peaks averaged around \SI{8}{GeV/c}, \SI{11}{GeV/c} and \SI{18}{GeV/c} that can be seen on the right of the central momenta peaks of \SI{6}{GeV/c}, \SI{8}{GeV/c} and \SI{12}{GeV/c} in \autoref{fig-spectrum}.}
    \label{fig-highmomentapions}
\end{figure}

The higher momentum particles observed in \autoref{fig-spectrum} especially in the \PiP\ beam are a consequence of the wider horizontal phase space and the particular configuration chosen for the collimator apertures. This collimator is \SI{5}{\meter} ($\approx30$ interaction lengths) long to ensure the complete absorption of the primary proton beam as well as a satisfactory absorption of the secondaries produced by the primary beam interaction with it. For this reason, the collimating slits of the \PiP\ and \PiM\ beams are effectively much larger than what strictly required by the $\mathcal{O}($25\%) theoretical momentum acceptance ($\SI{10.25}{\centi\meter}$ radius compared to $\SI{3.43}{\centi\meter}$ radius). Indeed, while the beam has a waist around the slit position, the size of the accepted beam at its entrance and exit is larger (see \autoref{fig-pions1dpp}) and therefore the momentum selection is less stringent. Furthermore, the wide horizontal angular acceptance for the \PiP\ beam in combination with the large aperture holes, allows for some \PiP's with higher momenta and large angles to be transported through the beamline, as shown in \autoref{fig-highmomentapions}. The overall content of the high-momentum tail is about 20\% of the total, and has been quantified by comparing the histogram integral in two different ranges: $[0,16]~\si{\giga\electronvolt}$/c and  $[16,30]~\si{\giga\electronvolt}$/c.

\begin{figure}
\centering
    \begin{subfigure}{0.49\textwidth}
    \includegraphics[width=\textwidth]{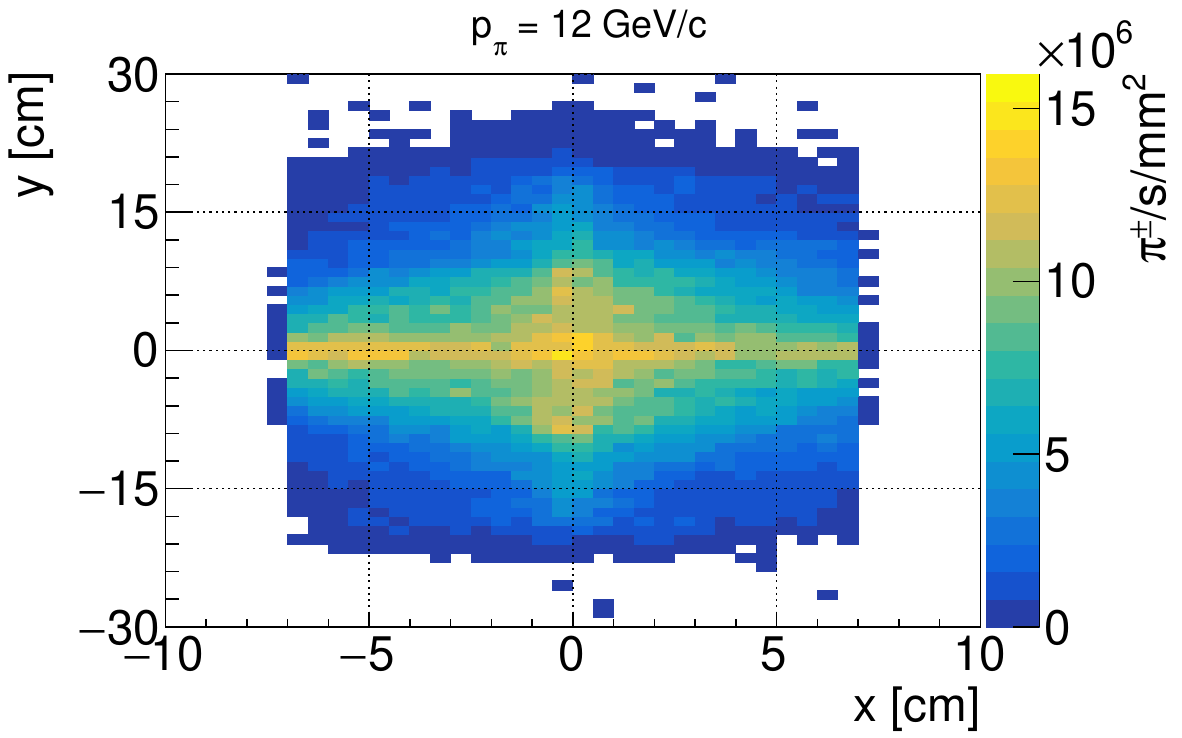}
    \caption{}
    \label{fig-beamsize12}
    \end{subfigure} \\
    \begin{subfigure}{0.49\textwidth}
    \includegraphics[width=\textwidth]{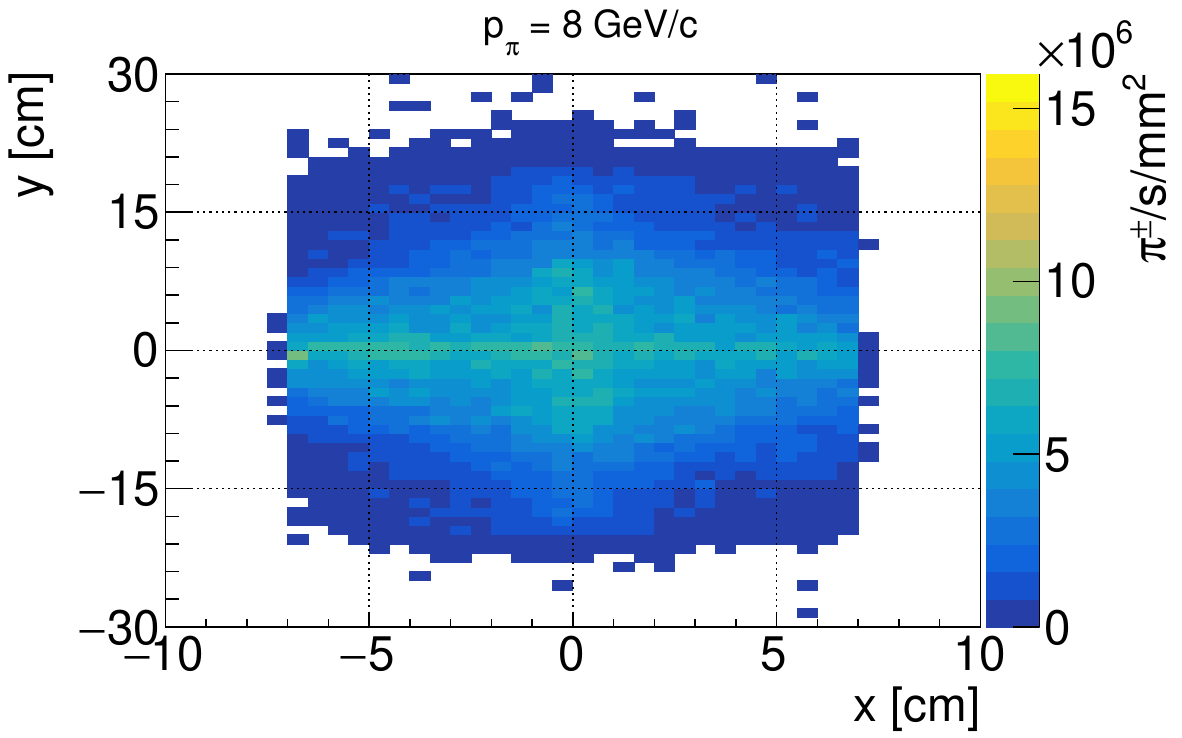}
    \caption{}
    \label{fig-beamsize8}
    \end{subfigure} \\    
    \begin{subfigure}{0.49\textwidth}
    \includegraphics[width=\textwidth]{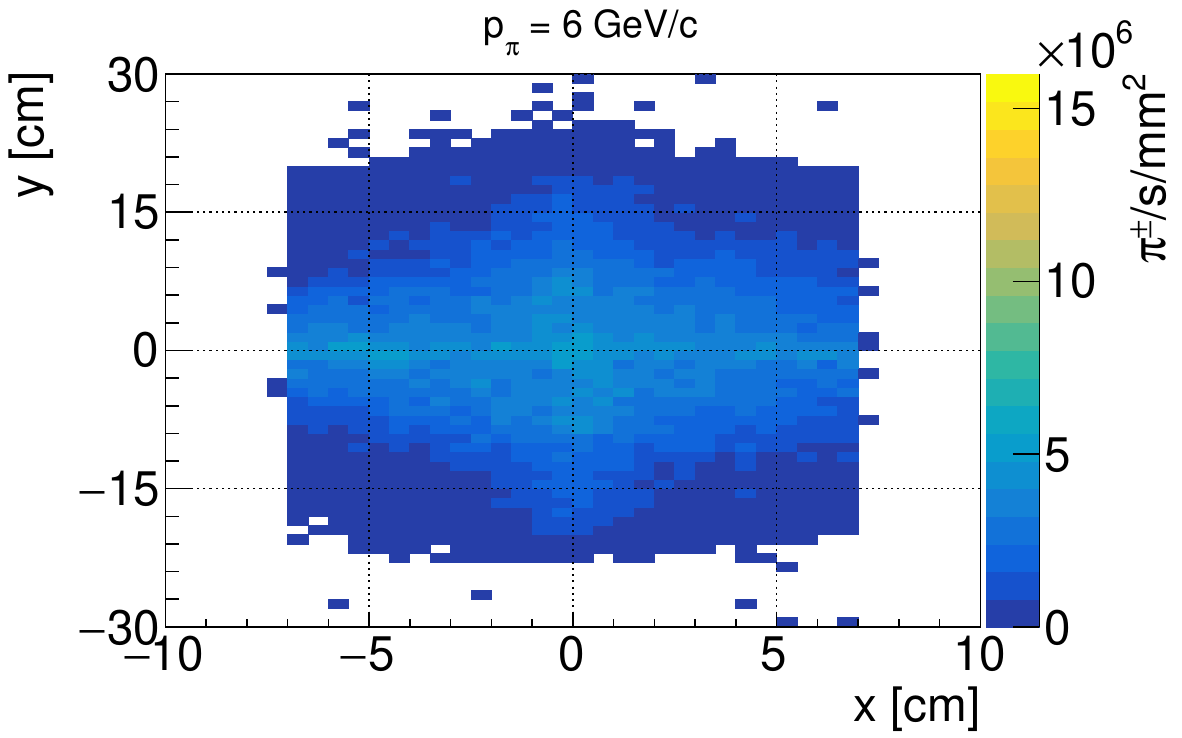}
    \caption{}
    \label{fig-beamsize6}
    \end{subfigure}  
\caption{$(x,y)$ map of the \Pipm\ flux after the second achromat ($z=$ \SI{63}{\metre}) obtained from the FLUKA simulation for a beamline with a nominal \Pipm\ momentum of \SI{12}{GeV/c} (a), \SI{8}{GeV/c} (b) and \SI{6}{GeV/c} (c), considering an impinging primary proton beam of $\SI{2.5\times10^{13}}{protons/spill}$ with a spill length of $\SI{4.8}{\second}$.}
\label{fig:beam_size}
\end{figure}

The simulation also allows to estimate the beam transverse size and particle flux at the second achromat. As explained in Section~\ref{sec:tracker}, the limitations of the tracker technology require the beam size to be \SI{<0.1}{m^2} and the rate \SI{10-100}{MHz/mm^2}. \autoref{fig:beam_size} shows a two-dimensional map of the beam particle flux at the exit of the second achromat. The two specifications are fulfilled with a beam size of about \SI{15\times50}{cm^2} and a peak flux of \SI{<15}{MHz/mm^2}, considering an impinging primary proton beam of $\SI{2.5\times10^{13}}{protons/spill}$ with a spill length of $\SI{4.8}{\second}$.

\subsection{Neutrino yield evaluation at the far detector}
\label{sec:piondecaykin}

Knowing the total number as well as the full 6D phase space ($x$, $x'$, $y$, $y'$, $\text{d}p/p$, $z$) of the pions transmitted through the beamline it is possible to precisely calculate the expected neutrino yield at the far detector. To do so, a custom calculation code has been developed.

\begin{figure}
\centering
    \begin{subfigure}{0.49\textwidth}
    \includegraphics[width=\textwidth]{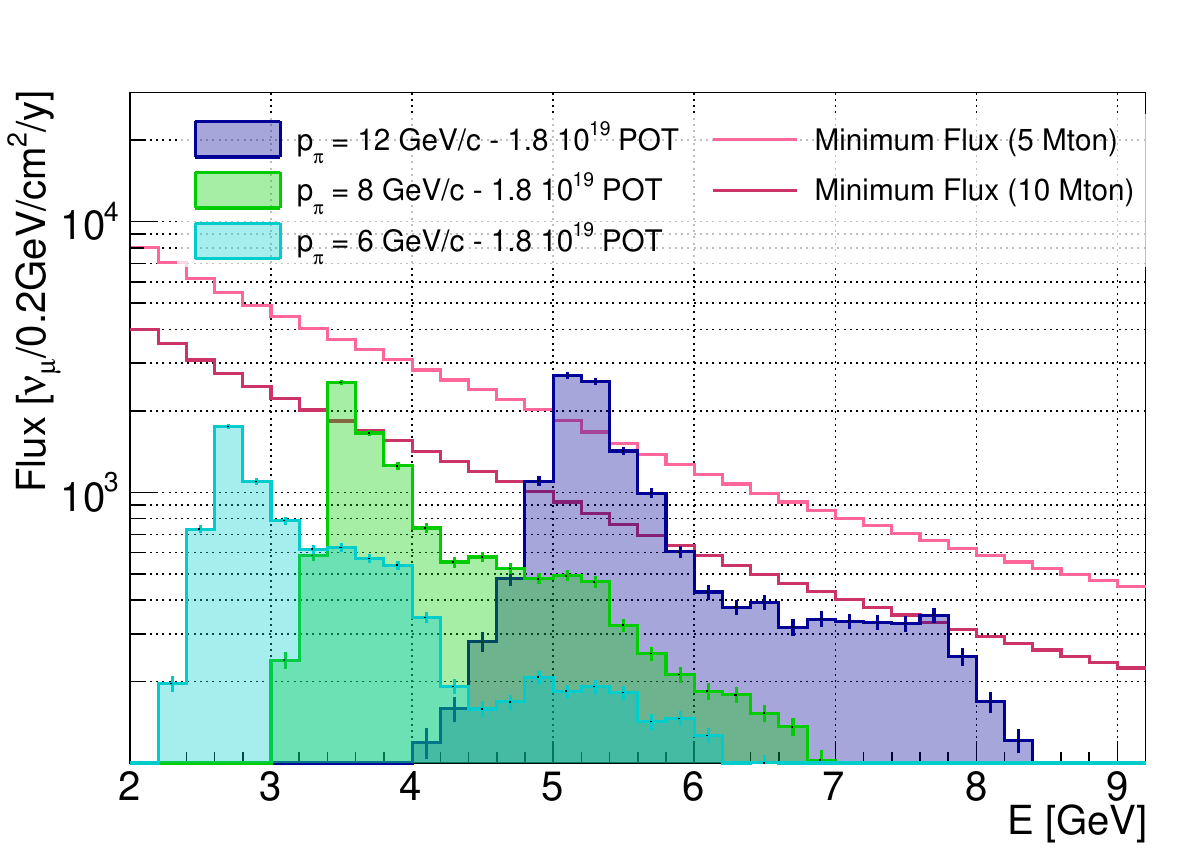}
    \caption{}
    \label{fig-YieldNu}
    \end{subfigure} \\
    \hfill
    \begin{subfigure}{0.49\textwidth}
    \includegraphics[width=\textwidth]{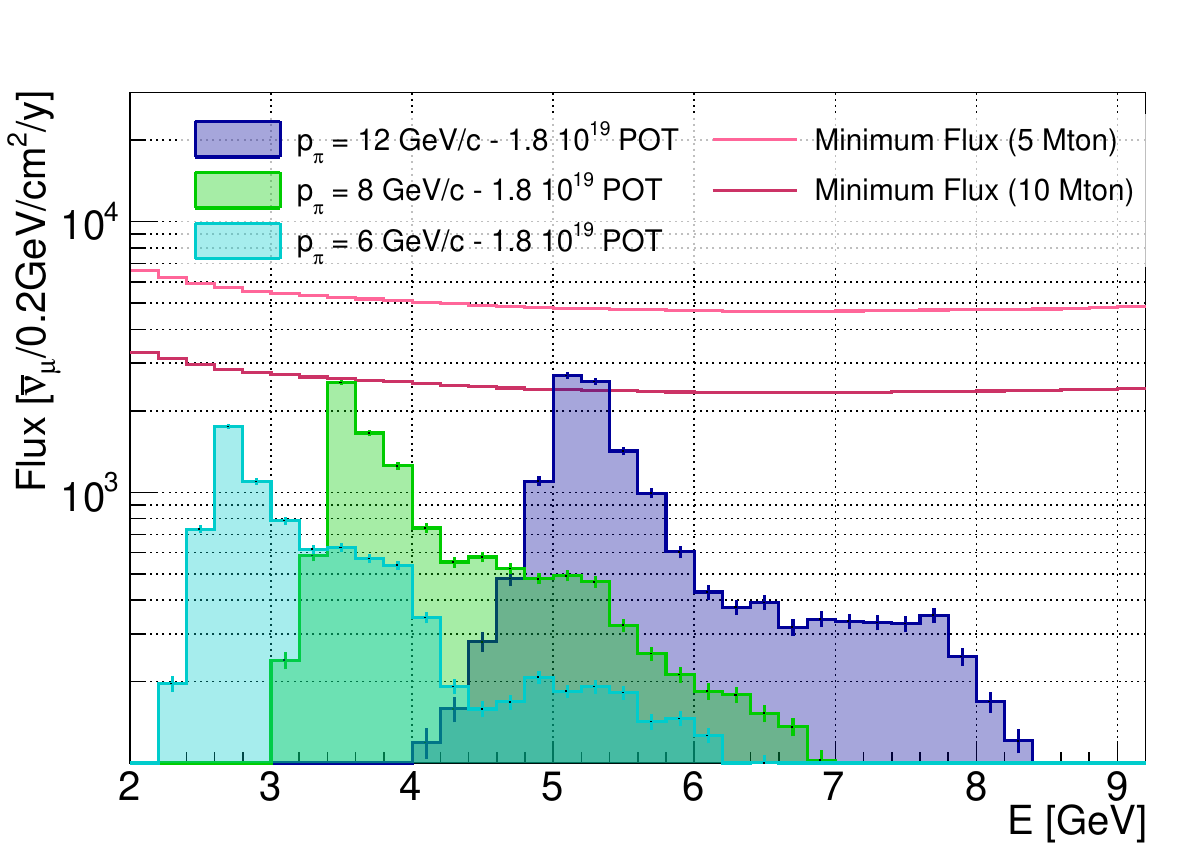}
    \caption{}
    \label{fig-YieldANu}
    \end{subfigure}      
\caption{\NuMu (a) and \ANuMu (b) fluxes from pion decays in the far detector acceptance for beamlines with nominal energies of \SI{12}{GeV} (dark blue), \SI{8}{GeV} (green) and \SI{6}{GeV} (light blue) overlaid with the minimal neutrino fluxes required to collect as many detected neutrino as DUNE~\cite{hep-ph_DUNE_2020} at the first oscillation maximum for a \SI{5}{Mton} (light red) and a \SI{10}{Mton} (dark red) detectors.}
\label{fig:neutrinoyields}
\end{figure}

The software, developed in C++, generates \Pipm's at the exit of the second achromat by sampling, for \PiP\ and \PiM\ separately, the $x$, $y$ two-dimensional distribution and $p$, $x'$, $y'$ three-dimensional distribution obtained from the FLUKA simulation at this position. For each \Pipm, the $z$ coordinate of a decay vertex, $z_{\rm{vtx}}$, is generated on an 
exponential distribution with a decay constant of $1/(\beta\gamma c\tau)$ where $\beta$ and $\gamma$ are the reduced velocity and Lorentz boost of the decaying \Pipm, $c$ the speed of light and $\tau$ the \Pipm\ mean lifetime~\cite{PhysRevD.98.030001}. Each \Pipm\ is then propagated to this $z_{\rm{vtx}}$ where the \PiMuNu\ is generated using standard decay kinematics~\cite{hep-ph_James_1968}. The resulting \NuANuMu\ are then propagated to the far detector, which is assumed to have a transverse half-size of \SI{100}{m}. The neutrino energy distribution at the far detector is obtained from the distribution of the generated \NuANuMu's which occur to be in the far detector acceptance. The flux per POT is obtained by scaling the previous distribution by the ratio of \Pipm\ per POT at the second achromat exit (obtained from FLUKA) and the number of generated \Pipm. The annual fluxes are shown in \autoref{fig:neutrinoyields} assuming $1.8\times10^{19}$ POT per year. This number corresponds to $2.5\times10^{13}$ protons per spill, 3000 spills per day, 30 days per month over the course of 8 months in a year of operation. In such conditions, the radiation resistance of the silicon pixel detector (see Sec~\ref{sec:tracker} would allow the experiment to operate for about 10 years.

The beamline is thus able to deliver a neutrino flux which would allow to collect, with megaton scale natural water Cherenkov detectors, neutrino samples with a size comparable to the ones of the upcoming LBL experiments. While systematical uncertainties are expected to be a limiting factor for the latter, they will be marginal for the former due to the tagging technique. Hence, this new generation of tagged LBL experiments, exploiting the beamline developed in this study, will be able to improve the knowledge on the neutrino oscillation in a sustainable way. Indeed, it requires beam powers of \SI{(10-100)}{kW} which, first, are much lower than the $\mathcal{O}(1)\unit{MW}$ beams of the upcoming experiments and of most of the proposed options for the next generation of LBL neutrino experiments~\cite{hep-ph_AlekouEtAl_2022,hep-ph_BogaczEtAl_2022} and, second, are partially available in existing facilities (however, not directly to the required intensity of $1.8\times10^{19}$ POT per year).

\section{Summary and Discussion}
\label{sec:conclusion}

The work presented here aimed to introduce a novel beamline concept for neutrino beam long baseline experiments, allowing to implement the neutrino tagging technique. This technique relies on the operation of charged particles spectrometers instrumented with trackers and installed in the beamline to kinematically reconstruct the \NuANuMu\ from \PiMuNu\ and \KMuNu\ decays. The operation of these trackers limits the beam particle flux and size. Indeed, at the tracker location, the charged beam particle flux has to be \SI{10-100}{MHz/mm^2} and its transverse size \SI{<0.1}{m^2}, constraints that results in a lower neutrino flux, compared to standard LBL neutrino experiments. This low neutrino flux is compensated by the use of \SI{\mathcal{O}(1)}{Mton} natural water Cherenkov detectors which are two orders of magnitude larger than the ones used in standard LBL neutrino experiments but allow for much lower precision on the neutrino characteristics. In a tagged experiment, this lower precision is compensated by the one of the kinematical reconstruction, that allows to estimate the neutrino energy with $<1$\% resolution, and to determine the neutrinos initial flavour and chirality.

The beamline proposed in this work allows to simultaneously transport \PiP and \PiM beams, halving the required run time to collect neutrino and anti-neutrino data while reducing the systematical uncertainties for key measurements such as the charge-parity violation phase, \deltacp, where the oscillation probabilities of neutrinos and anti-neutrino are compared. To fulfill the constraints on the particle flux, the beamline was designed to operate with a slow extracted proton beam. The study was performed assuming a \SI{400}{GeV/c} proton beam and, for this momentum, a graphite target of $\approx\SI{1.2}{m}$ was found to yield the largest amount of secondary \Pipm's. In a tagged beamline the neutrino flux is limited by the sustainable charged particle flux at the spectrometer which must be \SI{<100}{MHz/mm^2}. In the hereby proposed beamline, this flux is only \SI{20}{MHz/mm^2} per $\SI{2.5\times 10^{13}}{protons/spill}$ over a spill of \SI{4.8}{\second}. Hence, a significant margin is available to further improve the \Pipm\ yield.

With our study and the proposed tagged beamline, we demonstrated that it is possible to collect neutrino interaction samples with a size comparable to the one expected for upcoming experiments using beams of much lower power. As the tagging technique eliminates most of the systematical uncertainties limiting the sensitivity of these upcoming experiments, our tagged beamline offers a valuable option to further improve the knowledge on the neutrino oscillation parameters. At such tagged experiments, the precision on these parameters would then be limited by the size of the neutrino samples and thus by the neutrino fluxes. Therefore, any effort in the direction of improving the beamline performances should aim at maximizing the \Pipm yields and, consequently, the overall neutrino fluxes. This optimisation process would target two different objectives: (1) improvement of the target performances and (2) further improvement of the optics design to increase as much as possible the accepted phase space, including the acceptance through the final spectrometer. 

Particularly for the first part, the target geometry could be further optimized, including e.g segmentation that would possibly increase the pion yield. Also, a more complete optimisation process could be carried out, eventually considering other materials and graphite densities available in the market. This process should also include thermo-mechanical calculations to evaluate in depth the target station exact geometry and target life. 
Regarding the second point, various paths could be explored leading to an overall optimization of the optics and accepted phase space, in a similar manner as in~\cite{mussolini:ipac2021-thpab130}. As emerged from \autoref{fig-phasespacepionplus} and \autoref{fig-phasespacepionminus}, the vertical \SI{140}{\milli\meter} aperture of the bending magnets composing the first and second achromat, strongly limits the \Pipm\ accepted phase space. While a very strong field, and therefore small gap, is needed to separate the two charges at the momentum selection stage in the first achromat, the constraint on the aperture could be relaxed at the meson tracking station, where the resolution of the spectrometer is enough to measure \Pipm\ momentum with smaller displacements (for a $<1\%$ momentum resolution a minimum displacement from the beam axis of $\SI{5}{\centi\meter}$ is required, compared to the $\approx\SI{12}{\centi\meter}$ considered in this study). This would most likely improve also the \Pipm\ beam size and divergence, without exceeding the limits of the tagger maximum accepted rate. 
Another element affecting the phase space and momentum acceptance is the collimator. Besides momentum selecting the \Pipm\ beams, the purpose of this beamline element is to act as a dump for the primary proton beam. Our conceptual dump (comprising approximately 30 nuclear interaction lengths) is very effective in absorbing the primary beam and in cleaning the produced low energy secondaries from being transported downstream, however, other designs (e.g with asymmetric or tapered apertures) and a shorter length could be foreseen, finding the best compromise in the signal to background ratio.  

Finally, the analysis of the background and its implications on the performances of the tagger have not been treated in the current study but are fundamental to have a complete picture of the proposed beamline capabilities. Some preliminary simulations have been done to estimate the main sources of background, which are mainly muons produced by the interaction of the primary proton beam with the dump and neutrinos produced both in the target and in the dump. While the background neutrinos are mostly emitted at large angles and are not likely to reach the far detector, the muon background could be concerning both for the tagger, in terms of maximum charged particle rate, and for the neutrino yield measured at the far detector. Preliminary results show a muon background of $\approx \SI{0.015}{muons/primary}$ in the momentum range $p<\SI{14}{GeV/c}$ for a central momentum of \SI{12}{GeV/c}, compared to the $\approx$ 0.053 pions/primary \Pipm\ signal. Further studies will be conducted in the future to better understand the effects of this background on the overall performance of the beamline and on the precision on the measurement of the neutrino flux. In this context, inherent Monte Carlo uncertainties will also have to be looked at in more detail, including a respective analysis of fitting benchmark data. 

\bibliographystyle{apa}
\bibliography{main}

\end{document}